\documentclass[twocolumn,english,superscriptaddress,prl,aps]{revtex4}
\usepackage[T1]{fontenc}
\usepackage[latin9]{inputenc}
\usepackage{graphicx}
\usepackage{babel}
\usepackage{color}

\newcommand{\Litwo}{Li$_2$CuO$_2$}

\newcommand{\Livcu}{LiVCuO$_{4}$}

\begin{document}

\title{Interplay of inter-chain interactions and exchange anisotropy: \\
Stability of multipolar states in quasi-1D quantum helimagnets}

\author{S.\ Nishimoto}
\author{S.-L.\ Drechsler}
\thanks{Corresponding author: s.l.drechsler@ifw-dresden.de}

\affiliation{IFW-Dresden, P.O.~Box 270116, D-01171 Dresden, Germany}

\author{R.\ Kuzian}
\affiliation{IFW-Dresden, P.O.~Box 270116, D-01171 Dresden, Germany}
\affiliation{Institute for Problems of Materials Science NASU, Krzhizhanovskogo
3, 03180 Kiev, Ukraine}

\author{J.\ Richter}
\affiliation{Universit\"{a}t Magdeburg, Institut f\"{u}r Theoretische Physik, Germany}

\author{Jeroen van den Brink}
\affiliation{IFW-Dresden, P.O.~Box 270116, D-01171 Dresden, Germany}

\date{\today}


\begin{abstract}
We quantify the instability towards the formation of multipolar states in coupled spin-1/2 chain systems with a 
frustrating $J_{1}$-$J_{2}$ exchange,
in parameter regimes that are of directly relevance to edge-shared cuprate spin-chain compounds.
Three representative types of inter-chain coupling 
and the presence of uniaxial exchange anisotropy 
are considered. The magnetic phase diagrams are determined by Density Matrix Renormalization 
Group calculations 
and 
completed by exact analytic results for the nematic and dipolar phases. We establish that 
the residual couplings strongly affect the pitch of 
spiral states and 
their instability 
to multipolar phases. Our theoretical results bring to the fore 
novel candidate materials 
close to 
quantum nematic/triatic ordering. 
\end{abstract}

\maketitle

In a system
with frustrated magnetic interactions entirely new ground 
states (GS) can emerge from the ensuing competition. The geometric frustration 
of classical Ising spins on a pyrochlore lattice, for instance, results 
in the famous spin-ice state, the excitations of which are magnetic 
monopoles~\cite{Castelnovo08}. 
In 
frustrated
quantum magnets 
equally
exotic 
states such as spin liquids,
valence-bond crystals or nematic phases, can occur~\cite{FruMag11}.
In 
quantum spin chain systems, 
in particular, the competition between short and longer-range magnetic 
couplings is  a common source of frustration, a canonical example of which is the 
$J_1$-$J_2$ spin-1/2
chain~\cite{FruMag11}. 
Having antiferromagnetic (AFM) next nearest-neighbor (NNN) interactions 
($J_2>0$), it is frustrated for 
any sign of the nearest-neighbor (NN) coupling 
($J_1$). 
In the classical $J_1$-$J_2$ spin chain the competing interactions generate a helimagnetic state 
but in a single chain quantum fluctuations destroy the long-range helical order for any 
value of $J_1$.
For sufficiently high magnetic field, for any value of $J_1$, the FM state takes over and the system's 
magnons, its propagating spin-flips, become its exact single-particle excitations. 
The exchange parameters $J_1$ and $J_2$ determine the magnon dispersion and, in particular, 
the {\em interaction} between them. An AFM interaction leads to a repulsion 
between magnons, whereas a FM interaction results in an attraction,
which favors the formation of magnon bound states.
%
For a frustration ratio $\alpha$=$-J_{2}/J_{1} > 0.367$ an interesting and intensely studied nematic state can occur, which 
may be thought of as a condensate of 2-magnon bound 
states~\cite{Chubukov91,Kecke07,Vekua07,Hikihara08,Sudan09,Dmitriev09,Zhitomirsky10,Svistov11,Syromyatnikov12,Sizanov13}
characterized by a quadrupole spin order with a non-zero
anomalous average $\langle \hat{S}_i^{+}\hat{S}_j^{+}\rangle$.
For $1/4 <\alpha < 0.367$  also 3-, 4- and even higher magnon bound states can
condense, resulting in a rich phase diagram
with quite a number of exotic 
magnetic multipolar phases (MPPs). 

\begin{figure}
 \includegraphics[width=0.75\columnwidth]{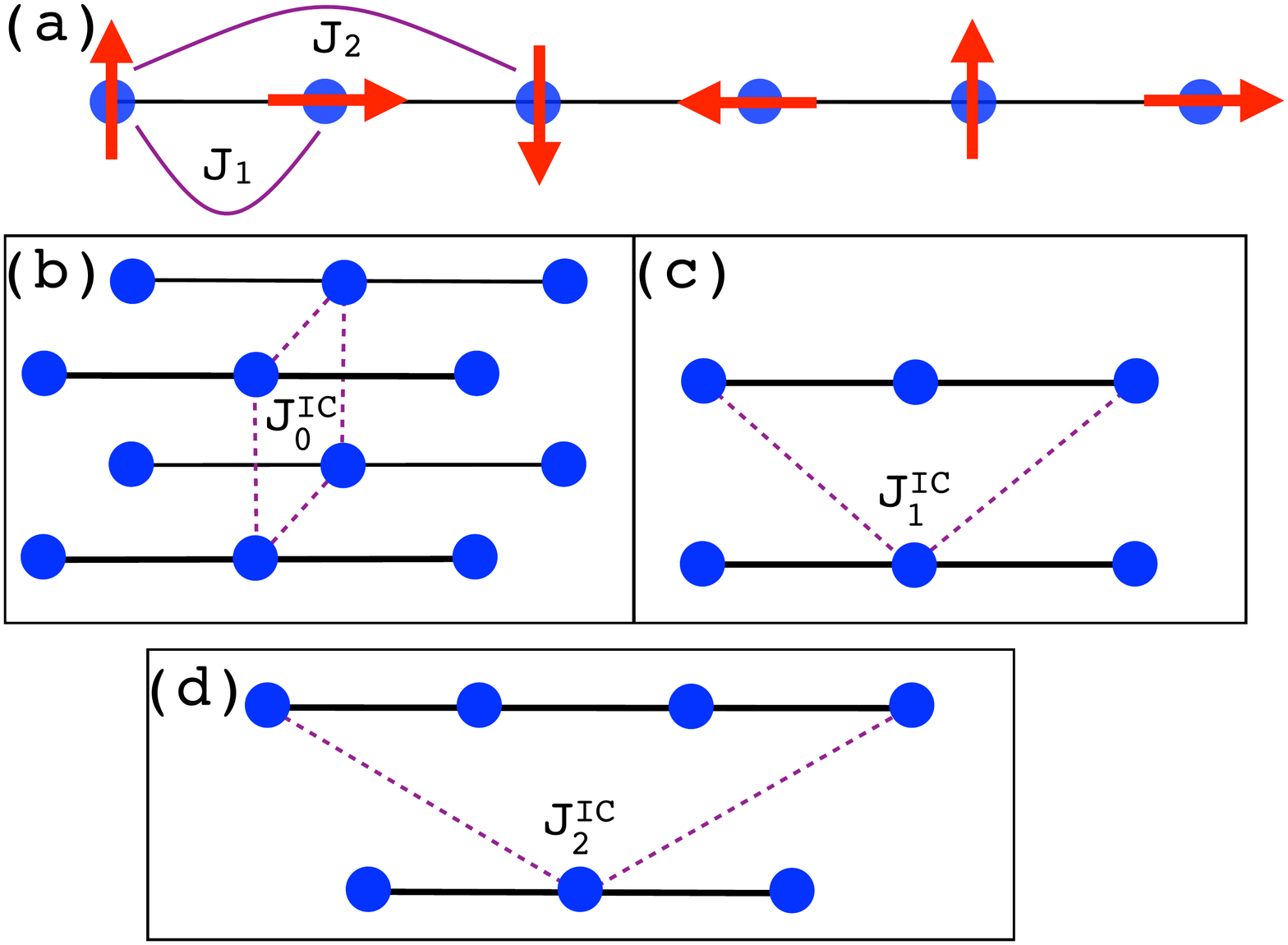}
 \caption{(a) Competing NN and NNN exchange $J_1$ 
 and $J_2$, respectively, along a chain. Coupling between different 
 chains: (b) perpendicular coupling $J^{\rm IC}_0$ (e.g., LiVCuO$_4$), (c) 
 skew (diagonal) coupling $J^{IC}_1$ (e.g., PbCuSO$_4$(OH)$_2$) and 
(d) skew NNN coupling between shifted chains $J^{\rm IC}_2$  (e.g., Li$_2$CuO$_2$).  
The effect of $J^{\rm IC}$ is considered in both 2D and 3D.}
\label{fig:fig1} 
\end{figure}

\textcolor{black}{
These theoretical developments have stimulated an experimental quest to 
find multipolar condensates in quasi one-dimensional (1D) magnetic materials, 
in particular in spin $s=1/2$ systems consisting of edge-sharing copper-oxide 
chains, such as LiVCuO$_4$ (in cuprate notation $\equiv$LiCuVO$_4$ in traditional chemical notation))~\cite{Enderle05,Buttgen07,Buttgen10,Hagiwara11,Svistov11}, 
Li$_2$ZrCuO$_4$~\cite{Drechsler07,Schmitt09}, Ca$_2$Y$_2$Cu$_5$O$_{10}$~\cite{Matsuda01,Kuzian12}, 
PbCuSO$_4$(OH)$_2$~\cite{Wolter12,Willenberg12}, Rb$_2$Cu$_2$Mo$_3$O$_{12}$~\cite{Hase04} 
and Li$_2$CuO$_2$~\cite{Lorenz09,Nishimoto11}. In these systems  $J_1$ 
is intrinsically FM and $J_2$ can be of comparable strength, but AFM. 
In real 3D materials, however, a magnetic inter-chain (IC) 
interaction is unavoidably present. Due 
to the fragility of purely 1D bound-states,  AFM IC interactions 
can pose a very relevant perturbation to a multipolar state, even when 
the coupling strength is (very) small~\cite{Nishimoto11}. 
To establish the consequences of this key ingredient for the stability 
of MPPs we consider here the three most common types of 
IC couplings $J^{\rm IC}$ that are encountered in the quasi-1D 
edge-shared cuprates mentioned above (one perpendicular IC coupling and 
two different types of skew ones, see Fig.\ \ref{fig:fig1}) and determine 
the boundaries of the magnetic phase diagram numerically by Density Matrix 
Renormalization Group (DMRG) calculations and analytically by} 
hard-core boson (HCB) \cite{Kuzian07,SM,Nishimoto11,Nishimoto12,Nishimoto12b})
\textcolor{black}{
and spin-wave (SW) \cite{Kuzian12} approaches. On top of this we consider also 
the presence of a uniaxial exchange anisotropy $\Delta -1$ for 
the NN coupling along the chains, which is the leading 
anisotropy term in edge-shared chain 
cuprates}~\cite{Tornow99,Yushankhai99,Kataev01,Heidrich09}. We show 
that the stability of MPPs is strongly affected by the strength of the
AFM IC couplings and depends on the precise 
type (geometry) of this coupling, which may also largely affect the pitch 
of the 
spiral state. A small easy-axis exchange anisotropy, however, enhances the stability 
of MPPs dramatically, also in the presence of IC coupling,
since it enhances the attraction between magnons.
From the material's viewpoint, our theoretical results bring to the fore 
linarite, PbCuSO$_4$(OH)$_2$, as a promising candidate compound with a  
triatic MPP,
which can be stabilized by its sizable exchange 
anisotropy and confirm the closeness of LiVCuO$_{4}$ to 
quantum nematicity.

The relevant Hamiltonian  $H  =  H_{1D}+ H_{IC}$ encompasses the frustrating 
magnetic interactions along the 1D chain in the presence of an external magnetic 
field $h$ and a small uniaxial exchange anisotropy $\Delta -1$
\begin{eqnarray}
 H_{1D} &= & \sum_{n,i} \left[-{\bf S}_{n,i} \cdot {\bf S}_{n,i+1} + \alpha  {\bf S}_{n,i}   \cdot {\bf S}_{n,i+2} \right] \nonumber \\ 
 &-&\sum_{n,i} \left[  (\Delta -1) S_{n,i}^z S_{n,i+1}^z   + hS_{n,i}^z  \right],
 \label{eq:H1DZ}
\end{eqnarray}
where $n$ labels the chain and $i$ the position of the spins along the chain. 
Neighboring chains $n$ and $m$ interact via 
\begin{eqnarray}
 H_{\rm IC} =  \sum_{\langle nm \rangle, r} J^{\rm IC}_r {\bf S}_{n,i} \cdot {\bf S}_{m,i+r}, 
 \end{eqnarray}
where $r=0$ corresponds to a perpendicular IC coupling and $r=1,2$ 
refer to skew
IC couplings, see Fig.\ \ref{fig:fig1}. We use 
$|J_1|$ as the energy unit of all coupling constants in $H$. 

To determine the nature of the magnetic GS and its dependence on the 
frustration $\alpha$, the different types of IC exchange $J^{\rm IC}$ 
and the exchange anisotropy $\Delta -1$, we employed the DMRG method~\cite{White92} 
with periodic boundary conditions (PBC) for all directions. This method 
is not restricted to purely 1D and can also be used for 
2D~ \cite{Nishimoto10,Stoudenmire12}
{\color{black}
 and  
3D~\cite{Nishimoto11,Nishimoto12b} systems, although the system size is 
limited, e.g.,  up to about $\sqrt{N} \times \sqrt{N} \times L = \sqrt{10} 
\times \sqrt{10} \times 50$ for spin Hamiltonians. We kept 
$p \approx 800-5000$ density-matrix eigenstates in the renormalization procedure. 
About $100-300$ sweeps are necessary to obtain the GS energy 
within a convergence of  $10^{-7}J_1$ for each $p$ value. All calculated 
quantities were extrapolated to $p \to \infty$ and the maximum error in 
the GS energy is estimated as $\Delta E /J_1 \sim 10^{-4}$, while the 
discarded weight is less than  $1 \times 10^{-6}$. Under the PBC, a 
uniform distribution of  $\left\langle S^z_i \right\rangle$ may give an 
indication to examine  the accuracy of DMRG calculations for spin systems. 
Typically,  $\left\langle S^z \right\rangle-S_{\rm tot}^z/(NL)$ is less 
than $1 \times 10^{-3}$  in our calculations. Note that for high-spin states  
[$S_{\rm tot}^z \gtrsim (NL-10)/2)$] the GS energy can be obtained with an 
accuracy of $\Delta E/J_1 < 10^{-12}$ by carrying out several thousands  
sweeps even with $p \approx 100-800$. 

We considered systems with different lengths: $L=16-64$ ($24-96$) for 3D  
(2D) and adopted power laws to perform a 
finite-size-scaling analysis. 
From this we obtained the saturation field $h_s$ in the thermodynamic 
limit $L\to\infty$. As a result, we obtain $h_s$ with high accuracy. In 
addition to DMRG we have also applied an analytic}  
HCB-approach 
and the
linear SW approach~\cite{Kuzian07,SM} to provide exact results for 
the nematic and dipolar phases. In addition, some of the calculated magnetization curves 
have been cross-checked by exact diagonalization.

{\color{black}
The simplest case, relevant for, e.g., LiVCuO$_4$ and  Li(Na)Cu$_2$O$_2$, 
is the situation of unshifted neighboring chains and a perpendicular 
inter-chain exchange $J^{\rm IC}_0$, see Fig.\ \ref{fig:fig1}a. In this 
case spirals on NN chains are only weakly affected by an AFM IC 
coupling~\cite{Zinke09} -- on a classical level the pitch of the 
incommensurate (INC) spiral state is not affected by $J^{\rm IC}_0$. This 
is in stark contrast to the effect of  skew AFM $J^{\rm IC}_1$ and 
$J^{\rm IC}_2$, which can strongly reduce the pitch. 
\begin{figure}[b]
 \includegraphics[width=.6\columnwidth]{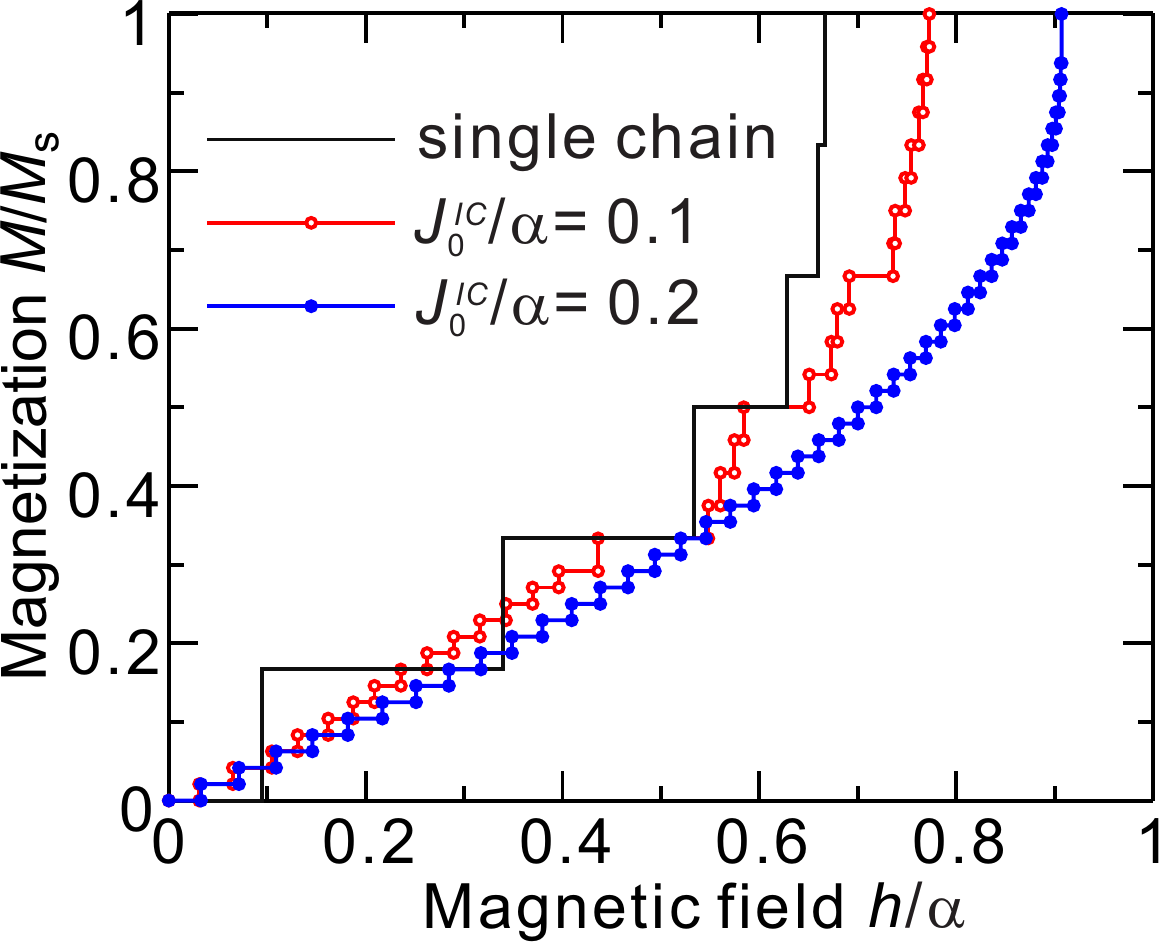}
\caption{
Magnetization vs.\ magnetic field for a 2D arrangement of four 
chains with  $N=24$ sites each, with a perpendicular IC coupling 
$J^{\rm IC}_0$ (cf.\ Fig.\ \ref{fig:fig1}b), $\alpha=1/2$ and $\Delta =1$. 
}
\label{fig:magnetization} 
\end{figure}
A typical magnetization curve for $\alpha$=0.5 and $\Delta =1$, 
for a nematic phase, is shown in Fig.\ \ref{fig:magnetization}. 
The height of the magnetization steps $\Delta S^{z}$=2 when 
{$J^{\rm IC}_0/\alpha$=0.1, 
is the direct signature for 2-magnon bound states. A larger value of 
the IC coupling suppressed these bound states, as is clear from the 
magnetization curve for $J^{IC}_0/\alpha$=0.2 where the steps correspond to 
$\Delta S^{z}$=1. So in the isotropic case, where $\Delta =1$,  a rather 
weak critical IC of a few percent destroys the nematic phase in favor of 
the usual conical ordering.  The critical value for 
$J^{IC}_0/(\alpha=0.5)$ amounts 0.188/0.088 in 2D/3D, respectively. The 
full phase diagram \cite{remark} is shown in Fig.\ \ref{fig:perp}, where 
the phase boundaries are extracted from the kinks in the calculated 
saturation field $h_{s}$ as a function of $J^{IC}_0$, as shown in 
Fig.\ \ref{fig:perp}(a-c). Clearly, the 3- , 4- , and higher multimagnon 
MPPs are even stronger affected by the IC interaction. 

Allowing for a finite uniaxial exchange anisotropy $\Delta -1$,  the 
leading-order anisotropy that is of immediate relevance to quasi-1D 
cuprates~\cite{Tornow99,Yushankhai99} affects the stability of the MPP 
substantially. Fig.\ \ref{fig:perp} shows that for $\alpha=1/2$ an 
anisotropy $\Delta -1$ of just 0.1 increases the critical IC coupling 
by a factor of $\sim$1.6, and thus significantly enhances their stability 
region.
\begin{figure}[b]
 \includegraphics[width=\columnwidth]{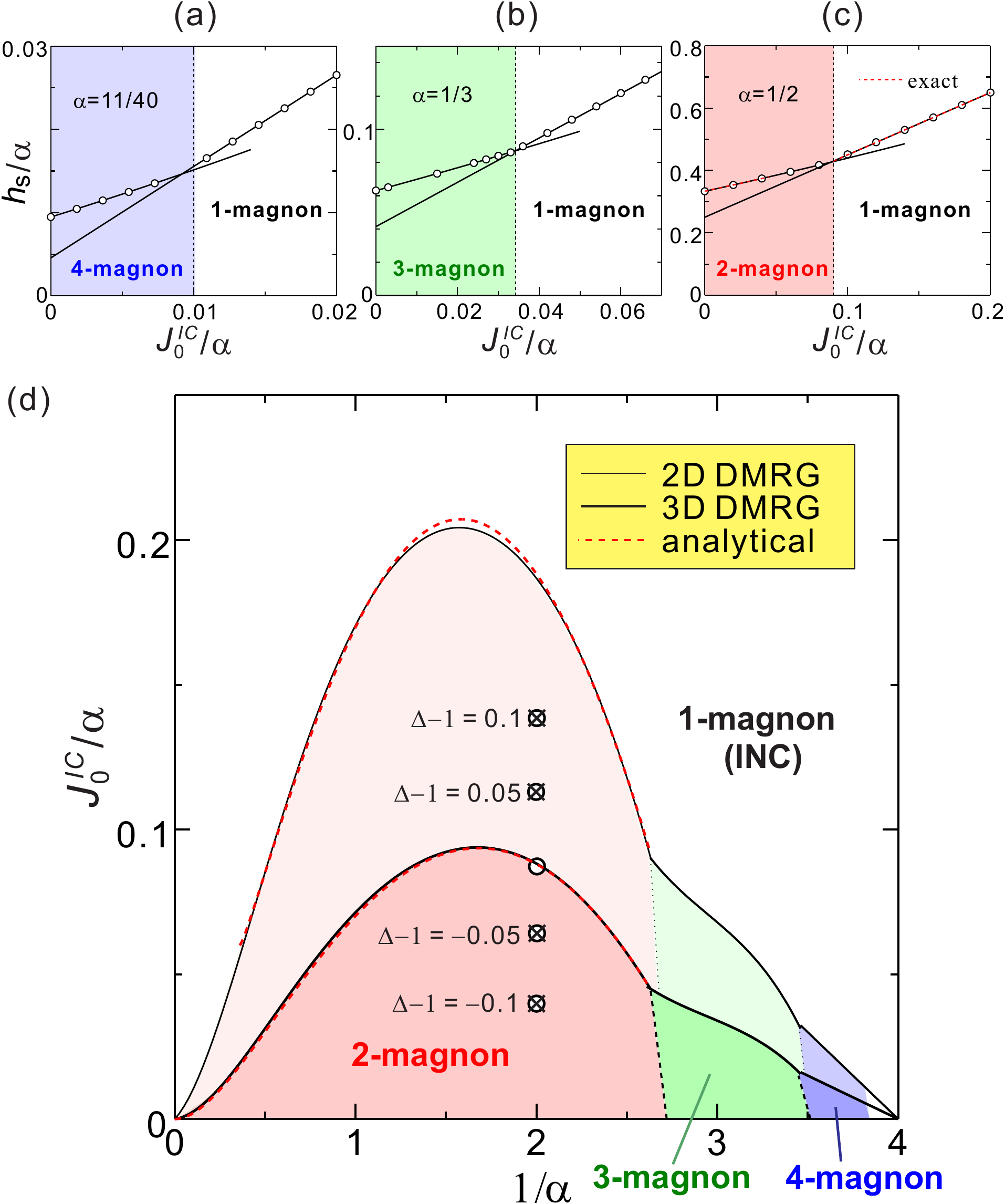}
\caption{
(a-c) Saturation field $h_s$ as a function of the perpendicular 
IC coupling $J^{\rm IC}_0$ (cf.\ Fig.\ \ref{fig:fig1}b) and $\Delta =1$.
(d) Phase diagram with critical IC coupling in 3D and 2D (thin line). 
Phase boundaries are extracted from the kinks in the $h_s$ as in (a-c). 
Red dashed lines: analytical HCB results [Eq.\ (\ref{eq:bet2})]. 
 Symbols: the dependence of the critical 
$J^{\rm IC}_0$ on the uniaxial exchange anisotropy $\Delta -1$
in 3D for $\alpha =0.5$, 
where $\circ$/$\times$ correspond to 
the DMRG/analytical HCB results,
respectively.
}
\label{fig:perp} 
\end{figure}

Our analytical approach to calculate the phase boundary between  the 
1- and 2-magnon instabilities relies on first deriving the saturation fields 
of these two instabilities: $h_{s,1}$ and $h_{s,2}$ respectively. Requiring 
them to be equal then renders the equation for the critical IC coupling 
as a function of anisotropy and frustration parameters. The saturation 
field $h_{s,1}$ of the INC phase on the 1-magnon side is 
exact
already within SW theory:
\begin{equation}
h_{s,1}= \frac{(4\alpha-1)^2}{8\alpha}+ \frac{N_{\rm IC}}{2} (J^{\rm IC}_0+|J^{\rm IC}_0 |)
- (\Delta -1) ,
\label{hs1mag}
\end{equation}
where $N_{\rm IC}$ denotes the number of  IC neighbors (i.e.\ for $J^{\rm IC}_0$ 
in 3D and 2D, $N_{\rm IC}=4$ and 2, respectively). In 
the Supplementary Material
this expression has been further generalized to 
include next NN and 
IC exchange anisotropies \cite{SM}. 
Therein we have shown
also that the critical value of $J_0^{\rm IC}$ of perpendicular IC
depends only on $\Delta$, $\alpha$, and $N_{\rm IC}$.
For the nematic phase we obtained \textit{exact} values of $h_s$ using 
the HCB-approach\cite{Kuzian07,SM}.
The HCB values 
{\color{black}
are in full accord 
with the DMRG results. In the limit 
$J^{\rm IC} \ll 1$ we arrive at the analytical expansion 
$
h_{\mbox{\tiny s,2}} \simeq  h_{\mbox{\tiny s}}^{\rm 1D}+\eta_{2}(J^{\rm IC})^2+ \eta_{4} (J^{\rm IC})^4,\label{hs2mag} 
$
which is approximate but accurate enough for our present purposes and where
$h_s^{\rm 1D} = -\Delta + 2\alpha+\Delta ^{2}/(2\Delta +2\alpha)$ 
\cite{Kuzian07},
and  $2 \eta_{2}(\alpha ,\Delta )=N_{IC}(\Delta +\alpha)(3\alpha^{2}+
3\alpha\Delta + \Delta ^{2})/[\Delta (\Delta +2\alpha)]^{2} 
\approx N_{IC}\left(5/6+3\alpha/4\right)$, when $\Delta \sim 1$. 
The expression for the next, quartic term $\eta_{4}$ is provided in 
Ref.\ \onlinecite{SM}. Comparing the expressions for $h_{s,1}$ and $h_{s,2}$ one notices the 
presence of nonlinear IC terms and a two times smaller linear term in 
the nematic phase as compared to the usual 1-magnon phase. The solution 
of the equation $h_{s,2}=h_{s,1}$ gives 
{ \it analytical} expressions for the 
critical IC interaction $J^{\rm IC}_{\rm 0,cr}$. Keeping only the
linear term in  
the expression for $h_{s,2}$, we find (cf.\ Eq.\ (51) in 
Ref.\ \onlinecite{Syromyatnikov12}) 
$
|J^{\rm IC}_{\rm 0,cr1}|= (4\alpha\Delta ^{2}-\Delta -\alpha)/
[4\alpha\left(\Delta + \alpha\right)N_{\rm IC}]$
and including the quadratic term~\cite{SM}, we obtain
\begin{equation}
|J^{IC}_{0,cr2}|=\frac{1}{4\eta_{2}}\left(N_{IC}-\sqrt{N_{IC}^{2}- 
8\eta_{2}N_{IC}|J^{IC}_{0,cr1}|}\right).
\label{eq:bet2}
\end{equation}
A comparison of the numerical DMRG results in Fig.\ \ref{fig:perp}  (cf.\ 
Fig.\ 6 of Ref.\ \onlinecite{Ueda09}) shows that 
 Eq.\ (\ref{eq:bet2})
is very accurate for 3D systems and works well for 2D ones, too.

\begin{figure}
 \includegraphics[width=\columnwidth]{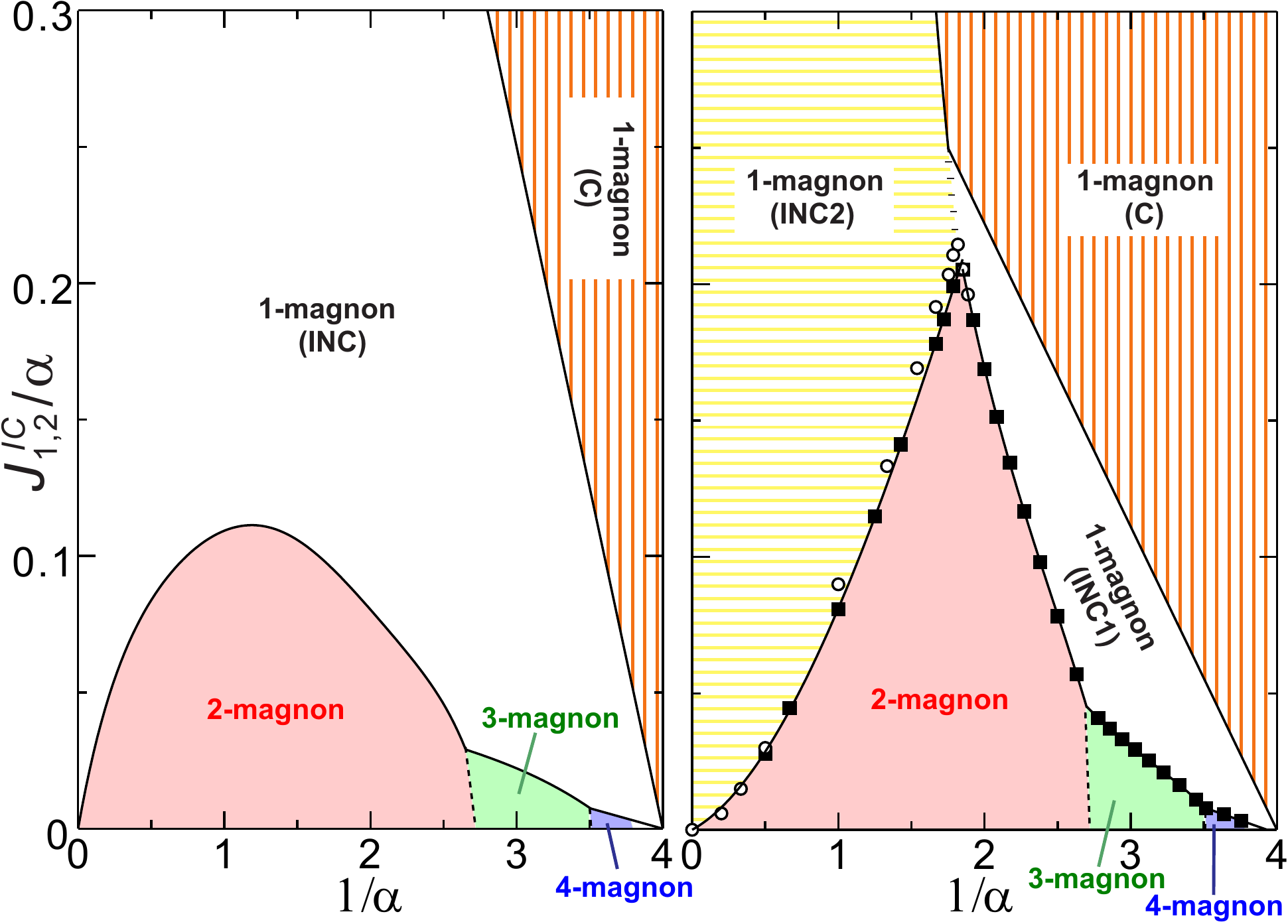}
\caption{Phase diagram for MPP swith skew (diagonal) 
IC coupling $J^{\rm IC}_1$ (left) and $J^{\rm IC}_2$ (right) in 3D. 
}
\label{f4} 
\end{figure}

The phase diagram for the situation of the two other, skew
types of IC interaction, $J^{\rm IC}_1$ and $J^{\rm IC}_2$ (see Fig.\ \ref{fig:fig1}c and d) 
are shown in Fig.\ \ref{f4}. An inspection of the phase diagrams reveals that 
the maximal value for the critical $J^{\rm IC}$ always occurs in the nematic 
phase at $\alpha$ slightly below 1, i.e.\ in the region of maximal in-chain 
frustration and quantum behavior~\cite{Kuzian07,Nishimoto12}. 
For the situation of perpendicular coupling this can be understood  
already in linear approximation, where $j_{\rm cr,1}$ is proportional to the 
difference of  1- and 2-magnon critical fields of an isolated chain 
$j_{\rm cr,1}=2(h_{\mbox{\tiny s}}^{\mbox{\tiny 1D}}- h_{s,1}^{\mbox{\tiny 1D}})/N_{\rm IC}$. 
Near the critical point  ($\alpha \gtrsim$ 1/4) and for almost decoupled Heisenberg chains the 
saturation  field tends to the simple 1-magnon value and additional quantum effects vanish.

Having investigated theoretically in general how the competition between 
frustration, different types of IC coupling and exchange anisotropy plays 
out, we now apply these insights to identify candidate 
materials potentially displaying a quantum MPP.
\Litwo \ is near the critical point, having $\alpha \approx 0.33$ and a 
rather small $\Delta -1 \approx 0.01$~\cite{Lorenz09}. 
Its IC coupling $J^{\rm IC}_2$, however, is strong enough to even destabilize 
the spiral state and drives the chains FM.
Also Li$_2$ZrCuO$_4$ is close to the critical point ($\alpha \approx 0.3$~\cite{Drechsler07}) 
but in this case as well for any realistic IC interaction and reasonable 
value for $\Delta $, all higher MPP are unstable.  
The compounds Li(Na)Cu$_{2}$O$_{2}$ are away from the detrimental critical point but 
their IC coupling is too large ($J^{IC}\sim 0.5$ to 1 
\cite{Gippius04,Masuda05,Drechsler06}) to establish a nematic phase for 
the estimated, moderate, values of $\Delta $~\cite{Mihaly06}.

Instead \Livcu \ is a good material for a nematic phase, 
having a coupling between the chains that is characterized by a very weak 
$J^{\rm IC}_0$, which manifests itself in strong quantum fluctuations 
evidenced by a small ordered magnetic moment ($0.3 \mu _{\mbox{\tiny B}}$) 
at low temperature and the observation of a 2-spinon continuum in inelastic 
neutron scattering \cite{Enderle10}.
The weak $J^{\rm IC}$ is also in accord with the fact that its saturation field
is close to the value of the uncoupled 1D-chain given by 
$h_s^{\rm 1D}$~\cite{Drechsler11}. In addition, the estimated 
$\alpha \approx 0.75$~\cite{Nishimoto12,Drechsler11}, near the maximum of 
the critical $J^{\rm IC}_{0, {\rm cr}}(1/\alpha )$-curve is almost optimal 
for  a nematic phase to survive (see Fig.\ \ref{fig:perp}).

\begin{figure}
 \includegraphics[width=.6\columnwidth]{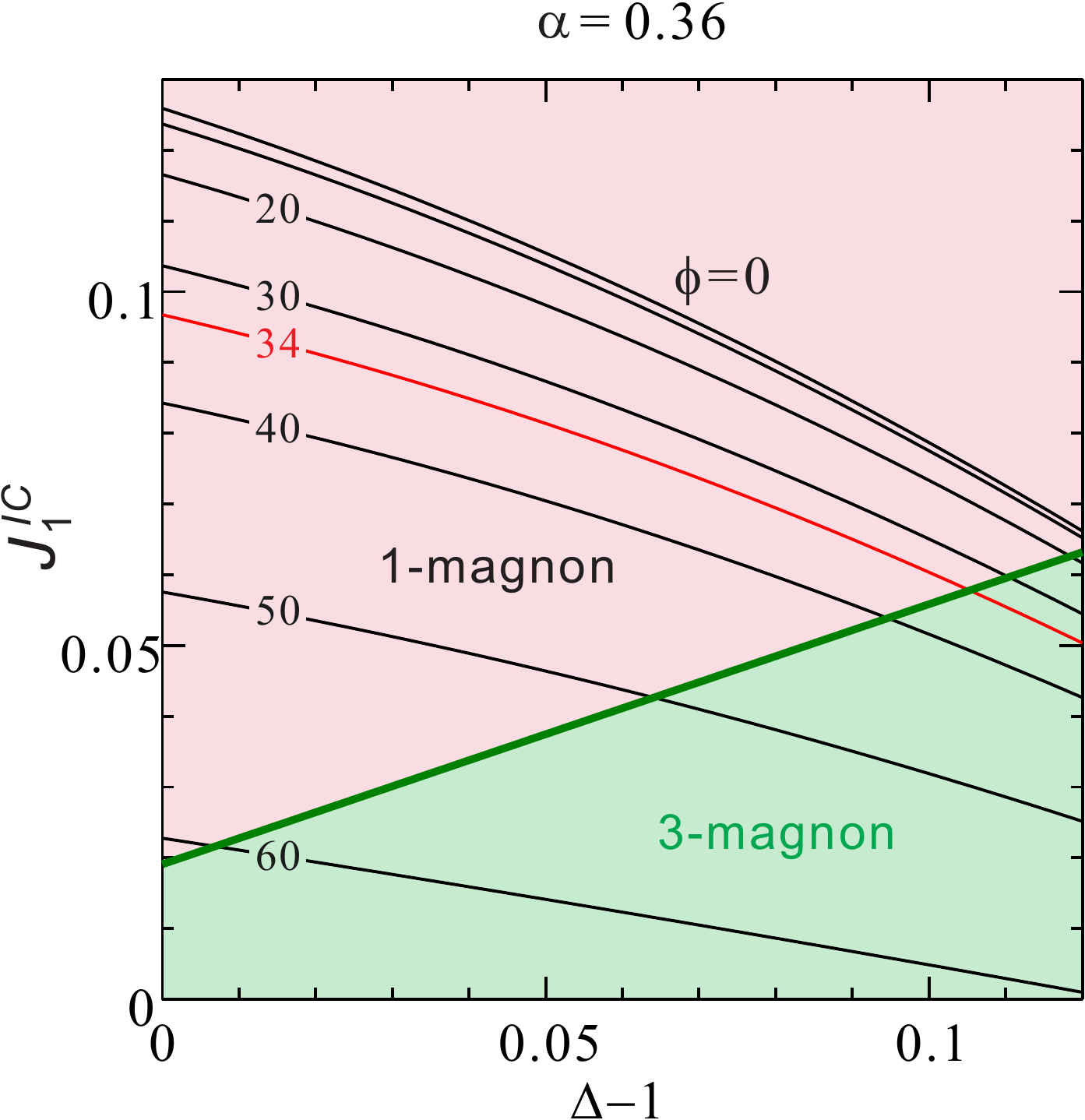}
\caption{Phase diagram and pitch  (contour lines) as a function of  
the diagonal IC coupling $J^{\rm IC}_1$ (in units of $|J_1|$)
 and the uniaxial exchange 
anisotropy $\Delta -1$ for  $\alpha=0.36$, as is relevant for linarite, 
PbCuSO$_4$(OH)$_2$. The red contour line corresponds to the experimental 
value of the pitch, 
34$^\circ$.
}
\label{fig:j_delta} 
\end{figure}

A very interesting case is provided by the natural mineral linarite, 
PbCuSO$_4$(OH)$_2$, which consists of neutral edge-shared Cu(OH)$_2$-chains
surrounded  by Pb$^{2+}$ and [SO$_4$]$^{-2}$ ions and has 
$\alpha \approx 0.36$~\cite{Wolter12}. Below 2.7~K a spiral state 
with a 
pitch 
of  34$^{\circ}$ sets in~\cite{Yasui11,Willenberg12}. 
A perpendicular $J^{\rm IC}_0$ barely affects the pitch 
of the 
spiral, in sharp contrast to skew 
IC coupling $J^{\rm IC}_1$. We have 
considered this situation theoretically in more detail and calculated the 
phase diagram as a function of $\Delta -1$ and $J^{\rm IC}_1$, see Fig.\ \ref{fig:j_delta}. 
For the given value of $\alpha$ a small $J^{IC}_1$ and $\Delta -1$ are enough to 
reduce the pitch from about 60$^{\circ}$ to the experimental value of 34$^{\circ}$. 
The
experimental 
pitch  
strongly restricts 
the possible values for 
$J^{\rm IC}_1$ and $\Delta -1$ (see the red line in Fig.\ \ref{fig:j_delta} ).
 An additional piece of information is the 
experimental value of the saturation field of 11 T
-- the 1D saturation 
field gives in this case about 5~T
-- which indicates a reduced value 
of  $J^{\rm IC}_1$, renormalized by a sizable $\Delta -1$, 
placing the system 
close to the triatic, 3-magnon region of the phase diagram in 
Fig.\ \ref{fig:j_delta}.

In this context experimental studies under chemical or physical pressure are of great
interest, since these can significantly change the IC coupling.
When applying hydrostatic pressure one expects an increase of the IC
coupling and thereby a weakening and possibly disappearance of the  MPPs in the mentioned 
two candidate materials.
Vice versa, growing
isomorphic crystals with larger isovalent
cations, i.e. substituting e.g.
Li or Na by 
Na, Rb, or Cs, respectively, is expected to lead to candidate MPP materials
due to a decrease of IC couplings. 
If possible to synthesize one expects e.g. for
Cs(Rb)Cu$_2$O$_2$ and Na(Rb)$_2$ZrCuO$_4$ an increased stability of the nematic
and triatic phase, respectively. Preparing strained epitaxial thin films from candidate materials will cause similar effects,
where a tuning of the strain can change the IC in different directions.  

We have, in summary, demonstrated the crucial role of different types of 
antiferromagnetic inter-chain interactions and the uniaxial exchange anisotropy 
in frustrated quasi-1D helimagnets. The rich and exotic physics of multipolar 
phases recently predicted for single chains is very sensitive to the strength 
and type and these additional and unavoidable interactions. 
Unfortunately, this prevents a  
realization of multipolar phases in most presently known 
spin-chain materials. But we find at least two 
notable exceptions:  LiVCuO$_{4}$, where a nematic phase is expected, and 
linarite, PbCuSO$_4$(OH)$_2$, which according
to our present calculations is in the close vicinity of a triatic instability.
In addition we proposed several new material systems as potential candidates with magnetic multipolar ground states and point out the large experimental potential of tuning the interchain interactions by pressure and strain.

We thank the DFG [grants DR269/3-1 (S-LD, SN), RI615/16-1 (JR)] for 
financial support and H.\ Rosner, A.\ Wolter, and M.\ Schaeper
for 
discussions on linarite.

\bibliographystyle{apsrev}
\bibliography{mpolar_rk2}

\begin{thebibliography}{47}
\bibitem[{rem()}]{remark}
\bibinfo{note}{Here and below, we use the presentation in terms of
  interpenetrating single Heisenberg chains coupled by $J_1$, which includes
  explicitely both the limit of two decoupled Heisenberg chains $1/\alpha$=0,
  and the quantum critical point $1/\alpha$=4 .}
\end{thebibliography}

\newpage

\widetext

\large
\begin{center}
{\bf Supplementary Material for\\
Interplay of interchain interactions and exchange anisotropy: \\
Stability of multipolar states in quasi-1D quantum helimagnets}

\normalsize
\vspace{3.0mm}
S.~Nishimoto$^1$, S.-L.~Drechsler$^1$, R.O.\ Kuzian$^{1,2}$, J.~Richter$^3$, Jeroen van den Brink$^1$\\

\small

\vspace{1.5mm}
{\it
$^1$IFW Dresden, P.O.~Box 270116, D-01171 Dresden, Germany\\
$^2$Institute for Problems of Materials Science NASU, Krzhizhanovskogo 3, 03180 Kiev, Ukraine\\
$^3$Universit\"at Magdeburg, Institut f\"ur Theoretische Physik, Germany\\
}

\small
\vspace{3.0mm}
\begin{minipage}{13.0cm}
We provide details on the derivation of the equations in the main
text, following the approach developed in Ref. 26.
The calculations are tedious but straightforward. 
\end{minipage}

\end{center}

\normalsize

\renewcommand{\theequation}{S\arabic{equation}}
\renewcommand{\thefigure}{S\arabic{figure}}
\renewcommand{\thetable}{S\arabic{table}}
\setcounter{equation}{0}
\setcounter{figure}{0}
\setcounter{table}{0}
\vspace{4.0mm}
At high magnetic fields, the Hamiltonian of coupled frustrated spin-1/2 chains
with the ferro- antiferromagnetic $J_{1}$-$J_{2}$  XXZ-Heisenberg model reads
\begin{eqnarray}
\hspace{-0.5cm} \hat{H} & = & \hat{H}_{1D}+\hat{H}_{ic}\label{H} \quad ,\\
\hspace{-0.5cm} \hat{H}_{1D} & = & \sum_{\mathbf{m}}\left[\frac{1}{2}
\sum_{\mathbf{r}}J_{\mathbf{r}}\left(\Delta_{\mathbf{r}}
\hat{S}_{\mathbf{m}}^{z}\hat{S}_{\mathbf{m}+\mathbf{r}}^{z}+
\hat{S}_{\mathbf{m}}^{+}\hat{S}_{\mathbf{m}+\mathbf{r}}^{-}\right)\right.\\
\hspace{-0.5cm}  &  & \left.-\mu \mathcal{H}\hat{S}_{\mathbf{m}}^{z}\right] \ , \label{eq:H1DZ}\\
\hat{H}_{ic} & = & \frac{1}{2}\sum_{\mathbf{f}}J_{\mathbf{f}}
\left[\Delta_{\mathbf{f}}\hat{S}_{\mathbf{m}}^{z}\hat{S}_{\mathbf{m}+
\mathbf{f}}^{z}+\hat{S}_{\mathbf{m}}^{+}\hat{S}_{\mathbf{m}+\mathbf{f}}^{-}\right] \ ,
\label{eq:Hperp}\end{eqnarray}
where $\mathbf{m}$ enumerates the lattice sites, ls
$\mathbf{r}=\pm n\mathbf{a},\: n=1, 2$
determines the NN sites within the chain, and $\mathbf{a}$ is 
the lattice vector along the chain. The vector $\mathbf{f}$ connects
sites at different chains. We restrict ourself to the case of 
uniaxial exchange anisotropy and the magnetic field directed along that 
axis, $\mu \equiv g\mu_{\mbox{\tiny B}}$.

In terms of hard-core boson operators $b$, defined by
\begin{equation}
\begin{array}{rcl}
\hat{S}^{+}\equiv b,\ \hat{S}^{-}\equiv b^{\dagger}, &  & \hat{S}^{z}\equiv\frac{1}{2}-\hat{n},\\
 &  & \hat{n}_{\mathbf{m}}=b_{\mathbf{m}}^{\dagger}b_{\mathbf{m}}=0,1,\\
\left\{ b_{\mathbf{m}},b_{\mathbf{m}}^{\dagger}\right\} =1, &  & 
\left[b_{\mathbf{m}},b_{\mathbf{m}^{\prime}}^{\dagger}\right]=0,\ \mathbf{m}\neq\mathbf{m}^{\prime},\ \\
b_{\mathbf{m}}^{\dagger}|FM\rangle & \equiv & b_{\mathbf{m}}^{\dagger}|
\cdots\uparrow\uparrow\uparrow_{\mathbf{m}}\uparrow\uparrow\uparrow\cdots\rangle\\
 & = & |\cdots\uparrow\uparrow\downarrow_{\mathbf{m}}\uparrow\uparrow\uparrow\cdots\rangle,\\
\left(b_{\mathbf{m}}^{\dagger}\right)^{2} & = & \left(b_{\mathbf{m}}\right)^{2}=0,
\end{array}
\label{eq:b}\end{equation}
the Hamiltonian (\ref{H}) becomes 
\begin{eqnarray}
\hat{H} & = & \hat{H}_{0}+\hat{H}_{int},\\
\hat{H}_{0} & = & \omega_{0}\sum_{\mathbf{m}}\hat{n}_{\mathbf{m}}+
\frac{1}{2}\sum_{\mathbf{m,R}}J_{\mathbf{R}}b_{\mathbf{m}}^{\dagger}b_{\mathbf{m+R}},\\
\hat{H}_{int} & = & \frac{1}{2}\sum_{\mathbf{m,R}}J_{\mathbf{R}}
\Delta_{\mathbf{R}}\hat{n}_{\mathbf{m}}\hat{n}_{\mathbf{m+R}},
\end{eqnarray}
where $\omega_{0}\equiv \mu \mathcal{H}-\frac{1}{2}\sum_{\mathbf{R}}
J_{\mathbf{R}}\Delta_{\mathbf{R}}$,
$\mathbf{R}=\mathbf{r,f}$.

The $n$-particle excitation spectra are given by the singularities of
the corresponding retarded Green's functions (GF) 
\begin{eqnarray}
\langle\langle\hat{X}|\hat{Y}\rangle\rangle & \equiv & -
\imath\int_{t^{\prime}}^{\infty}\!\! dte^{i\omega(t-t^{\prime})}\left\langle 
\left[\hat{X}(t),\hat{Y}(t^{\prime})\right]\right\rangle ,\\
\omega\langle\langle\hat{X}|\hat{Y}\rangle\rangle & = & 
\left\langle \left[\hat{X},\hat{Y}\right]\right\rangle +
\langle\langle\left[\hat{X},\hat{H}\right]|\hat{Y}\rangle\rangle .
\end{eqnarray}
A negative value of the excitation energy signals an instability of the
ground state, which is given by the fully polarized state achieved 
for a magnetic field above the saturation field 
$\mathcal{H}> \mathcal{H}_{s}$.

The equation of motion for the two-magnon operator 
\begin{equation}
\begin{array}{rll}
\hat{A}_{\mathbf{k,l}} & = & \frac{1}{\sqrt{N}}\sum_{\mathbf{m}}
\mathrm{e}^{-\imath\mathbf{k}(\mathbf{m}+\mathbf{l}/2)}b_{\mathbf{m}}b_{\mathbf{m+l}}=
\hat{A}_{\mathbf{k,-l}},\end{array}\label{eq:Akl}
\end{equation}
 reads
 \begin{equation}
\begin{array}{lll}
\left[\hat{A}_{\mathbf{k,l}},\hat{H}\right] & = & \left(2\omega_{0}+
\sum_{\mathbf{R}}J_{\mathbf{R}}\Delta_{\mathbf{R}}\delta_{\mathbf{l,R}}\right)\hat{A}_{\mathbf{k,l}}\\
 & + & \left(1-\delta_{\mathbf{l,0}}\right)\sum_{\mathbf{R}}
 J_{\mathbf{R}}\cos\frac{\mathbf{kR}}{2}\hat{A}_{\mathbf{k,l+R}},
 \end{array}\label{eq:AklH}\end{equation}
where $\mathbf{k}$ being the total quasi-momentum of the magnon pair, 
$N=N_{\perp}N_{x}$
is the number of sites, $N_{\perp}$ is the number of chains, and
$N_{x}$ denotes the number of sites in the chain. %
\begin{figure}
\includegraphics[width=0.4\columnwidth]{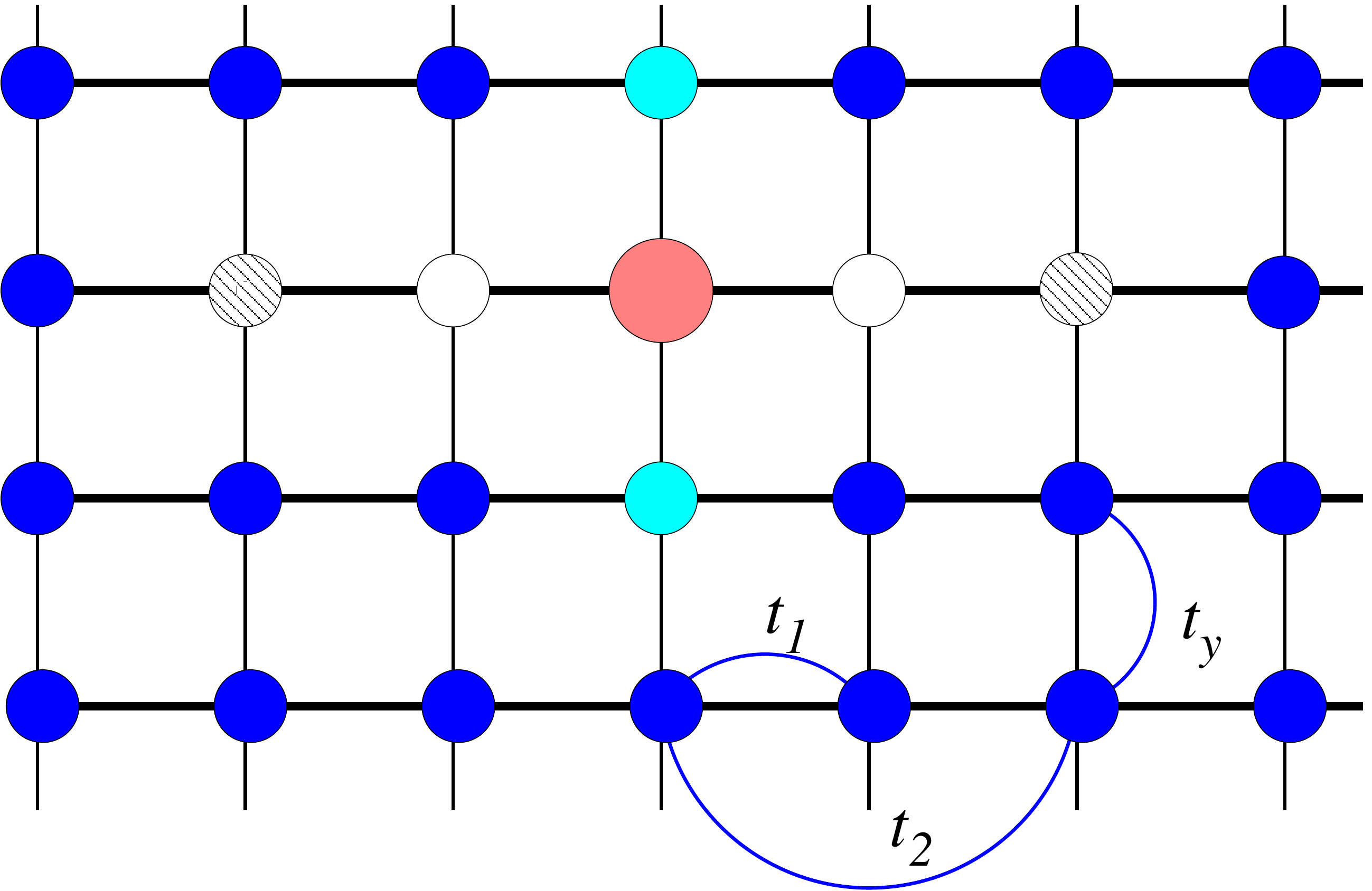} 
\caption{
Cartoon of the effective impurity problem given
by the Hamiltonian (\ref{eq:Htb}), which describes the internal motion
of a magnon pair with the total quasi-momentum $\mathbf{k}$. The
pink, open, shaded and cyan circles depict the impurities with 
$\varepsilon_{\mathbf{m}}
=\infty,J_{1}\Delta_{1},J_{2}\Delta_{2},J_{\perp}\Delta_{\perp}$
respectively, $\bullet $ : 
the regular sites of the lattice,
arcs: the $\mathbf {k}$-dependent hoppings. }
\label{fig1} 
\end{figure}

As usual, the exclusion of the center of mass motion reduces 
the problem of an interacting pair particles to a one-particle
problem of motion 
in an effective potential well (EPW). In our case it corresponds
to an impurity problem in a tight-binding Hamiltonian
\cite{Kuzian07}
(see Fig.\ \ref{fig1})
\begin{eqnarray}
\hat{H}_{tb}(\mathbf{k}) & = & \hat{T}(\mathbf{k})+\hat{V},\label{eq:Htb}\\
\hat{T}(\mathbf{k}) & = & 2\omega_{0}\sum_{\mathbf{m}}
\left|\mathbf{m}\right\rangle \left\langle \mathbf{m}\right|\label{eq:T}\\
 &  & +\sum_{\mathbf{m,R}}\left|\mathbf{m+R}\right\rangle 
 t_{\mathbf{R}}(\mathbf{k})\left\langle \mathbf{R}\right|,\\
\hat{V} & = & \sum_{\mathbf{m}^{\prime}}\left|\mathbf{m}^{\prime}\right\rangle \varepsilon_{\mathbf{m}^{\prime}}\left\langle \mathbf{m}^{\prime}\right|,
\label{eq:V}\end{eqnarray}
where \begin{eqnarray}
t_{\mathbf{R}}(\mathbf{k}) & = & J_{\mathbf{R}}\cos\frac{\mathbf{kR}}{2},\label{eq:t}\\
\mathbf{m}^{\prime}=\mathbf{0,r,f} &  & \varepsilon_{\mathbf{0}}=\infty,\ \varepsilon_{\mathbf{R}}=J_{\mathbf{R}}\Delta_{\mathbf{R}}.
\label{eq:impe}\end{eqnarray}
The Hamiltonian depends on the total pair momentum.

The two-magnon GF reads 
\begin{eqnarray}
G_{\mathbf{l,n}}(\mathbf{k},\omega) & = & \left\langle \left\langle A_{\mathbf{k,l}}|A_{\mathbf{k,n}}^{\dagger}\right\rangle \right\rangle ,\label{Gfi}\\
 & = & \left\langle \phi_{\mathbf{l}}\right|\left(\omega-\hat{H}_{tb}\right)^{-1}\left|\phi_{\mathbf{n}}\right\rangle \label{Gfitb}
\end{eqnarray}
 with $\left|\phi_{\mathbf{l}}\right\rangle =\left(\left|\mathbf{l}\right\rangle +\left|-\mathbf{l}\right\rangle \right)/\sqrt{2}$.
The GF  is analytic everywhere in the
complex energy plane but 
may have singularities on  the real axis: branch cuts 
and isolated poles. The branch cuts correspond 
to the continuum spectrum of 
unbounded motion of the effective particle, which in 
its turn correspond to
the two-particle continuum in the
pair motion. The poles correspond to the energies
of localized impurity states, which are bound states for the pair when 
the energies lie below the
continuum or anti-bound states in the opposite case.
It is clear from Eqs.~(\ref{eq:Htb})-(\ref{eq:impe}) that bound states 
are possible only when some $\varepsilon_{\mathbf{R}}$ are negative, 
i.e.\ for FM $J_{\mathbf{R}} < 0$. The bound state energy and the continuum
boundaries depend on the
total momentum of the pair $\mathbf{k}$. If the bound
state energy minimum lies below the lowest continuum energy (that may occur at different 
 $\mathbf{k}$-values), the bound pairs will condense at magnetic fields 
 just below the saturation field, the gas of pairs being the nematic state
of the magnetic system\cite{Chubukov91,Syromyatnikov12}. 

When all $J_{\mathbf{R}}$ are positive,
like in AFM-AFM $J_1$-$J_2$ model, only anti-bound states occur at energies 
higher the two-particle continuum. In this case only the one-magnon 
condensation occurs below the saturation field.

We will use the identity 
\begin{equation}
\hat{G}=\hat{g}+\hat{g}\hat{V}\hat{G} \ ,
\label{eq:Dy}\end{equation}
for the solution in the real space of the impurity problem
given by Eqs.\ (\ref{eq:Htb})-(\ref{Gfitb}) (see Fig.\ \ref{fig1}). 
In Eq.\ (\ref{eq:Dy}),
$\hat{g}\equiv\left(\omega-\hat{T}\right)^{-1}$ is the resolvent
operator for the periodic part, and $\hat{G}\equiv\left(\omega-\hat{H}_{tb}\right)^{-1}$
is the resolvent for the impurity problem. According to 
Ref.\ \onlinecite{Economou},
we may solve the problem step by step. Starting from the GF of a free
particle, which in the matrix form reads
\begin{eqnarray}
g_{\mathbf{l,n}} & = & g_{\mathbf{l-n}}(\mathbf{k},\omega)\label{eq:g}\\
 & = & \frac{1}{N}\sum_{\mathbf{q}}\frac{\cos\mathbf{q}(\mathbf{l-n})}{\omega-\left(\omega_{\mathbf{k}/2+\mathbf{q}}^{SW}+\omega_{\mathbf{k}/2-\mathbf{q}}^{SW}\right)},\\
\omega_{\mathbf{q}}^{SW} & = & \omega_{0}+\frac{1}{2}\sum_{\mathbf{R}}J_{\mathbf{R}}\mathrm{e}^{\imath\mathbf{qR}},
\label{eq:wSW}\end{eqnarray}
 we add the impurity at the origin. Its infinite potential reflects
the impossibility to have two particles on the same site (\ref{eq:b})
\begin{eqnarray}
g_{\mathbf{l,n}}^{(\mathbf{0})} & = & g_{\mathbf{l,n}}+
g_{\mathbf{l\mathbf{,0}}}\varepsilon_{\mathbf{0}}g_{\mathbf{0,n}}^{(\mathbf{0})},\nonumber \\
g_{\mathbf{l,n}}^{(\mathbf{0})} & = & g_{\mathbf{l,n}}+\frac{g_{\mathbf{l,0}}
\varepsilon_{\mathbf{0}}g_{\mathbf{0,n}}}{1-\varepsilon_{\mathbf{0}}g_{\mathbf{0,0}}}
\rightarrow g_{\mathbf{l,n}}-\frac{g_{l\mathbf{,0}}g_{\mathbf{0,n}}}{g_{\mathbf{0,0}}}.
\label{eq:g0ln}\end{eqnarray}
Next, we add an impurity at the site $\mathbf{i}$ and express the
GF via $\hat{g}^{(\mathbf{0})}$
\[
g_{\mathbf{l,n}}^{(i)}=g_{\mathbf{l,n}}^{(\mathbf{0})}+
\frac{g_{\mathbf{l,i}}^{(\mathbf{0})}\varepsilon_{\mathbf{i}}
g_{\mathbf{i,n}}^{(\mathbf{0})}}{1-\varepsilon_{\mathbf{i}}g_{\mathbf{i,i}}^{(\mathbf{0})}},
\]
 and so on, the GF of the system with $r$ impurities is expressed
via the GF of the system with $r-1$ impurities
\begin{equation}
g_{\mathbf{l,n}}^{(r)}=g_{\mathbf{l,n}}^{(r-1)}+\frac{g_{\mathbf{l,r}}^{(r-1)}
\varepsilon_{\mathbf{r}}g_{\mathbf{r,n}}^{(r-1)}}{1-\varepsilon_{\mathbf{r}}g_{\mathbf{r,r}}^{(r-1)}}.
\label{eq:grln}\end{equation}
 Thus, in principle, we may take into account any number of in-chain
and inter-chain exchange couplings (IC) and obtain $G_{\mathbf{l,n}}(\mathbf{k},\omega)$
(\ref{Gfi}). The explicit expression for the GF $G_{1,1}(k,\omega)$
for the 1D $J_{1}$-$J_{2}$ model (\ref{eq:H1DZ}) has been given in 
Ref.\ \onlinecite{Kuzian07}.
It's spectral density is plotted in Fig.\ \ref{fig2}. 
\begin{figure}
\includegraphics[width=.45\columnwidth]{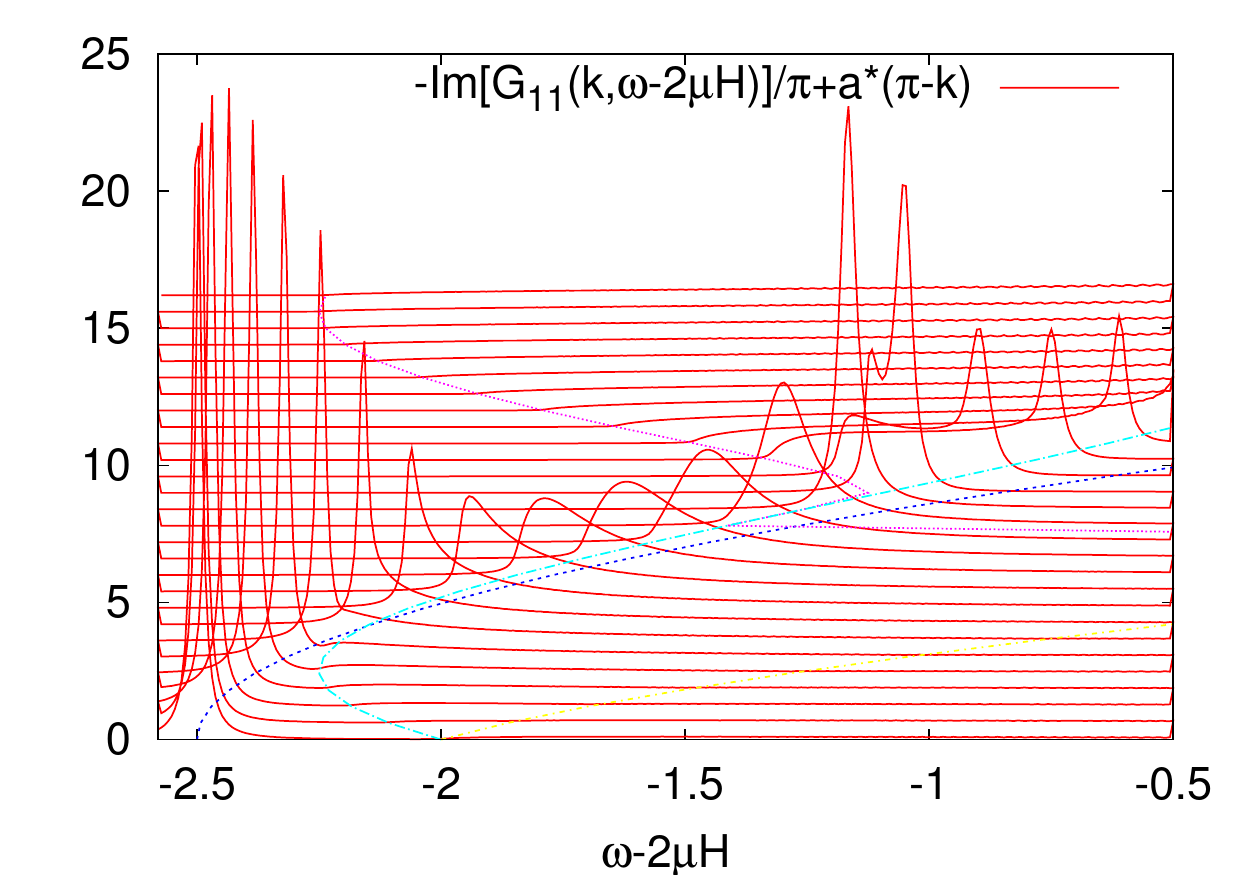} 
\caption{The spectral density of the two-particle Green's function for 
an isolated chain. 1D case, i.e. $J_{1}$=-1,\ $J_{2}$=1, $J_{\perp}$=0.
Cyan and magenta thin lines shows the lower boundary of the 2-magnon continuum.}
\label{fig2} 
\end{figure}
The sharp $\mathbf{k}$-dependent peaks below the two-particle continuum 
corresponds to bound pairs of magnons.

At higher dimensions, the role of the inter-chain interaction
(\ref{eq:Hperp}) is twofold. First, the periodic part of the effective
Hamiltonian (\ref{eq:T}) becomes D-dimensional. This changes $\hat{g}$
from Eq.\ (\ref{eq:g}) via the change of $\omega_{\mathbf{q}}^{SW}$ 
(\ref{eq:wSW}). Second, new impurities with the strength 
$\varepsilon_{\mathbf{r}}=J_{\perp}\Delta_{\perp}$
are added at points $\mathbf{r}$. The simplest geometry for the IC
corresponds to $\mathbf{f}$-vectors perpendicular to the chains,
which connect NN sites, only. The spectral density for GF
$G_{\mathbf{a,a}}(\mathbf{k},\omega)$ for $\mathbf{k}\parallel\mathbf{a}$
for the $2D$ case is depicted in Fig.\ \ref{fig3}.
\begin{figure*}
\includegraphics[width=0.45\textwidth]{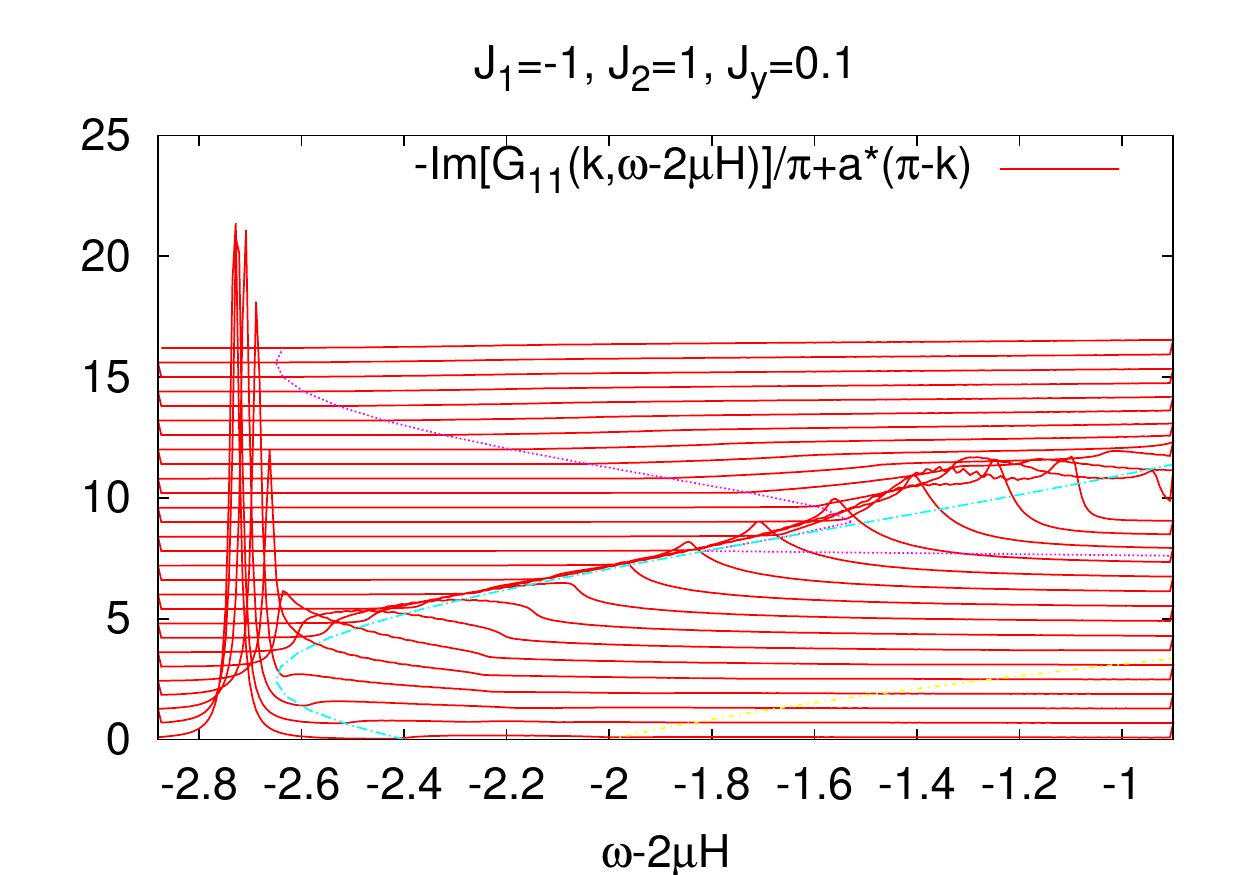} 
\includegraphics[width=0.45\textwidth]{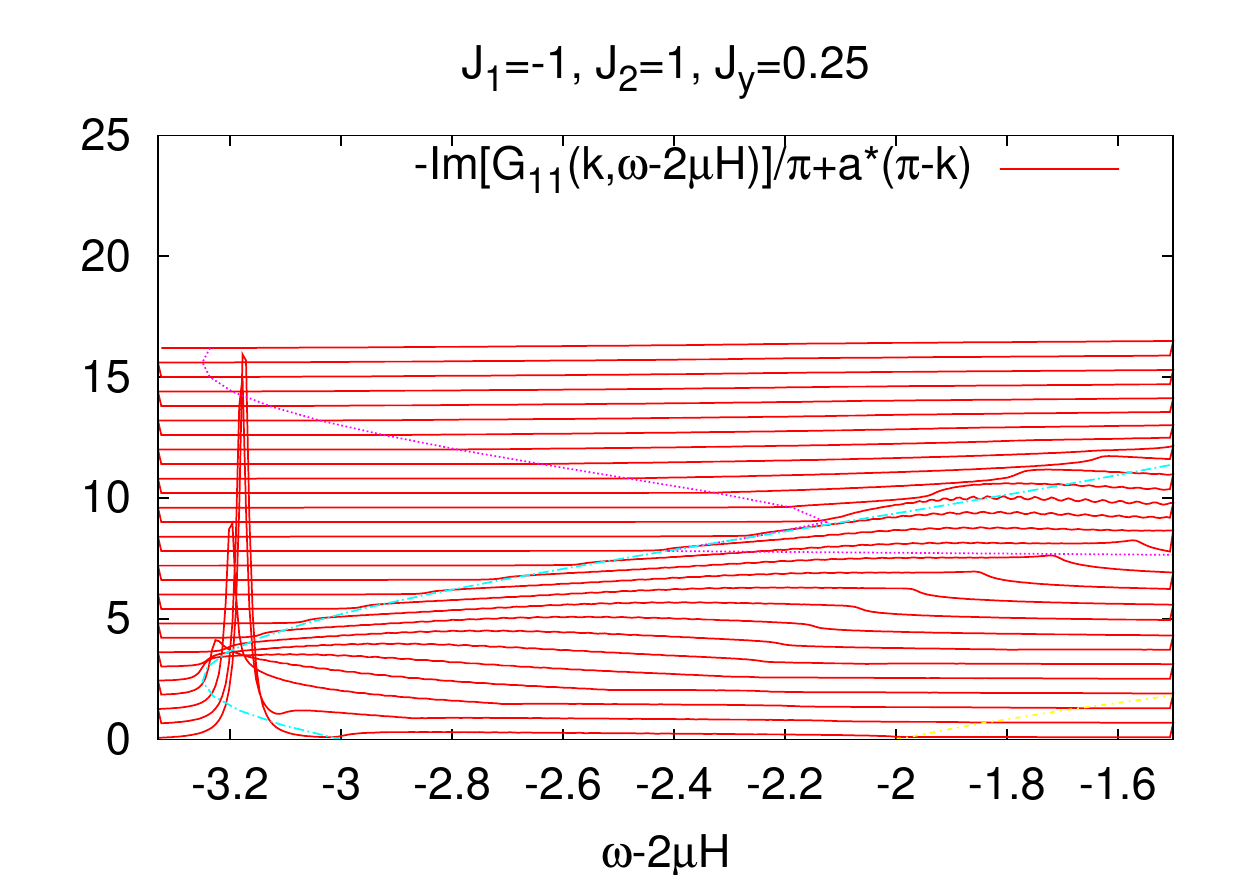}
\caption{The spectral density of the two-particle Green's function for
a 2D arrangement of unshifted $J_1$-$J_2$ chains and perpendicular IC 
interaction, left: $J_{y}\equiv J_{\perp}<J_{cr}$, 
right: $J_{y}>J_{cr}$. }
\label{fig3} 
\end{figure*}
We see that for small IC couplings the spectral density behaves qualitatively
similar to the 1D case, i.e.\ the peak corresponding to the bound pair
lies below the continuum (left panel of Fig.\ \ref{fig3}), 
and its dispersion exhibits a minimum at the total
momentum $\mathbf{ka}=\pi$ of a pair. We have checked numerically that
the minimum position remains at the point 
$\mathbf{k}_{\pi}=\left(\pi/a,0,0\right)$
for all values of IC satisfying the condition $J_{\perp}<J_{cr}$. On the right panel
of Fig.\ \ref{fig3} we see that the behavior of the spectral density changes
for large enough IC. The bound state is still present near the edge
of the Brillouin zone, but its energy is higher than the minimum of
the two-particle continuum. It is clear that the critical IC value
$J_{cr}$ is defined by the condition 
\begin{equation}
\omega_{b}\equiv\omega(\mathbf{k}_{\pi})=\omega_{min},
\end{equation}
 where 
$\omega_{min} = 2\left(\mu \mathcal{H}- |J_1|h_{s,1}\right)$
 is the minimum of the energy of the two-particle
continuum, and
\begin{equation}\begin{array}{rcl}
h_{s,1} & \equiv & 
-\Delta_{1}+\alpha\left(\Delta_{2}+1\right)+ \\
 &+& 0.125/\alpha +0.5N_{\mbox{\tiny ic}}\left(j_{\mbox{\tiny ic}}\Delta_{ic}+|j_{\mbox{\tiny ic}}|\right) \ ,
\end{array}\label{eq:hs1}\end{equation}
 is the critical field of the 1-magnon instability (Eq.\ (1) of the main text).
In order to find the expression for the saturation field $\mathcal{H}_{s}$
as a function of IC $\left|J_{\perp}\right|<J_{cr}$, we need the
expression for $\omega_{b}$, which is the position of an isolated pole
of the GF
\begin{equation}
\left[G_{\mathbf{a,a}}(\mathbf{k}_{\pi},\omega_{b})\right]^{-1}=0.
\label{eq:pole}
\end{equation}
In terms of the effective model $\hat{H}_{tb}(\mathbf{k}_{\pi})$
(\ref{eq:Htb}), $\omega_{b}$ is the energy of the localized impurity
level. From Eq.\ (\ref{eq:t}) we see that the nearest-neighbor hopping
along the chain vanishes $t_{\mathbf{a}}=J_{1}\cos\frac{\pi}{2}=0$,
and the sites with $\mathbf{r}=n\mathbf{a}+m\mathbf{b}+l\mathbf{c}$
having odd and even $n$'s are decoupled. In the subsystem with odd
$n$'s, only two impurities of the same strength $\varepsilon_{\mathbf{a}}=J_{1}$
are present at the sites $\pm\mathbf{a}=(\pm a,0,0)$. 
The effective particle
motion is {\emph not} affected neither by the
impurity at the origin (of infinite
strength) nor by the impurities at the sites $\mathbf{f}=(0,\pm b,0),(0,0,\pm c)$
with the energies $J_{2}\Delta _{2}$, and $J_{\perp}\Delta _{\perp}$, 
respectively. 
Note that this peculiarity has an important consequence: the critical 
value of the IC given below by 
Eqs.(\ref{eq:bet2})-(\ref{eq:y2}) {\em depends only} on the 
nearest-neighbor exchange anisotropy value $\Delta _1$. 
So, we may immediately 
write down the expression for the GF (cf.\ Eq.\ (49) of Ref.\ 
\onlinecite{Kuzian07}) 
\begin{equation}
G_{\mathbf{a,a}}(\mathbf{k}_{\pi},\omega)=
\left\{ \left[G_{\mathbf{a,a}}^{(0)}(\mathbf{k}_{\pi},\omega)\right]^{-1}-J_{1}\Delta_{1}\right\} ^{-1},
\label{eq:49pap}\end{equation}
 where\begin{eqnarray}
G_{\mathbf{a,a}}^{(0)}(\mathbf{k}_{\pi},\omega) & = & 
\left\langle \phi_{\mathbf{a}}\right|\left(\omega-\hat{T}(\mathbf{k_{\pi}})-
\left|\mathbf{0}\right\rangle \varepsilon_{\mathbf{0}}\left\langle 
\mathbf{0}\right|\right)^{-1}\left|\phi_{\mathbf{a}}\right\rangle \nonumber \\
 & = & g_{\mathbf{0}}(\mathbf{k_{\pi}})+g_{2\mathbf{a}}(\mathbf{k_{\pi}})-
 \frac{2g_{\mathbf{a}}^{2}(\mathbf{k_{\pi}})}{g_{\mathbf{0}}}\label{eq:G0kpigen} \\
 & = & g_{\mathbf{0}}(\mathbf{k_{\pi}})+g_{2\mathbf{a}}(\mathbf{k_{\pi}}).
 \label{eq:G0kpi}\end{eqnarray}
 In Eq.\ (\ref{eq:G0kpigen}) we have used the relation (\ref{eq:g0ln}) 
 and Eq.\ (\ref{eq:G0kpi})
follows from $g_{\mathbf{a}}(\mathbf{k_{\pi}})=0$, since the vector
$\mathbf{a}$ joins two decoupled subsystems. Then Eq.\ (\ref{eq:pole})
may be rewritten as
\begin{equation}
G_{\mathbf{a,a}}^{(0)}(\mathbf{k}_{\pi},\omega)=
\left(J_{1}\Delta_{1}\right)^{-1}
\label{eq:polepi}\end{equation}

Now, using the definition (\ref{eq:g}), we may write 
\begin{eqnarray}
\hspace{-1.0cm}G_{\mathbf{a,a}}^{(0)}(\mathbf{k}_{\pi},\omega) & = & \hspace{-0.2cm}
\frac{1}{N_{\perp}}\sum_{q_{y},q_{z}}G_{1,1}^{(0)}\left(\pi,\omega-E_{\perp}(\pi,\mathbf{q})\right),\label{eq:G0aa}\\
\hspace{-0.8cm}G_{1,1}^{(0)}\left(\pi,\omega\right) & = & \frac{1}{N_{x}}\sum_{q_{x}}\frac{1+\cos2q_{x}a}{\omega-E_{1D}(\pi,q_{x})},\label{eq:Gpi1D}\\
E_{1D}(\pi,q_{x}) & = & 2\left[\mu \mathcal{H}+J_{1}\left(\cos q_{x}a-\Delta_{1}\right)\right.\nonumber \\
\hspace{-1.0cm} &  & +\left.J_{2}\left(\cos2q_{x}a-\Delta_{2}\right)\right],\label{eq:E1D}\\
\hspace{-1.0cm}E_{\perp}(\pi,\mathbf{q}) & = & N_{ic}J_{\perp}\left(\gamma_{\mathbf{q}}-\Delta_{\perp}\right),
\label{eq:Eperp}\end{eqnarray}
 where $\gamma_{q}=\cos q_{y}b$ ($\left(\cos q_{y}b+\cos q_{z}c\right)/2$),
$N_{\rm ic}=2$(4) for a 2D (3D) geometry, respectively. In the 2D case the summation over
$q_{z}$ should be dropped. The 1D GF as given by Eq.\  (\ref{eq:Gpi1D})
is easily calculated 
\begin{eqnarray}
G_{1,1}^{(0)}\left(\pi,\omega\right) & = & G(z)/J_{2},\label{eq:G1Dz}\\
G(z) & = & \left[z+1-\tau(z)\right]^{-1},
\end{eqnarray}
 where we have introduced the dimensionless variable
\begin{equation}
z(\omega)  \equiv  \left[\omega-2\left(\mu \mathcal{H}-J_{1}\Delta_{1}-J_{2}\Delta_{2}\right)\right]/J_{2},
\end{equation}
 and the dimensionless Green's function of a semi-infinite tight-binding
chain $\tau(z)=\left[z-\tau(z)\right]^{-1}$. Now, we search for the
solution of Eq.\ (\ref{eq:49pap}) in the form 
\begin{equation}
\omega_{b}=J_{2}\left(z_{b1}+\zeta\right)+2\left(\mu \mathcal{H}-J_{1}\Delta_{1}-J_{2}\Delta_{2}-\frac{1}{2}N_{\rm ic}J_{\perp}\Delta_{\perp}\right),
\label{eq:wb}\end{equation}
 where $\zeta $ is unknown, and
 \begin{equation}
z_{b1}\equiv-\left(\frac{\Delta_{1}+\alpha}{\alpha}+\frac{\alpha}{\Delta_{1}+\alpha}\right) \quad ,
\label{eq:zb1}
\end{equation}
 is the solution for the 1D-problem \cite{Kuzian07}. Note that here we
use another definition for the frustration parameter $\alpha\equiv J_2/|J_{1}|$
as compared to Ref.\ \onlinecite{Kuzian07}. %
\begin{figure}
\includegraphics[width=.45\columnwidth]{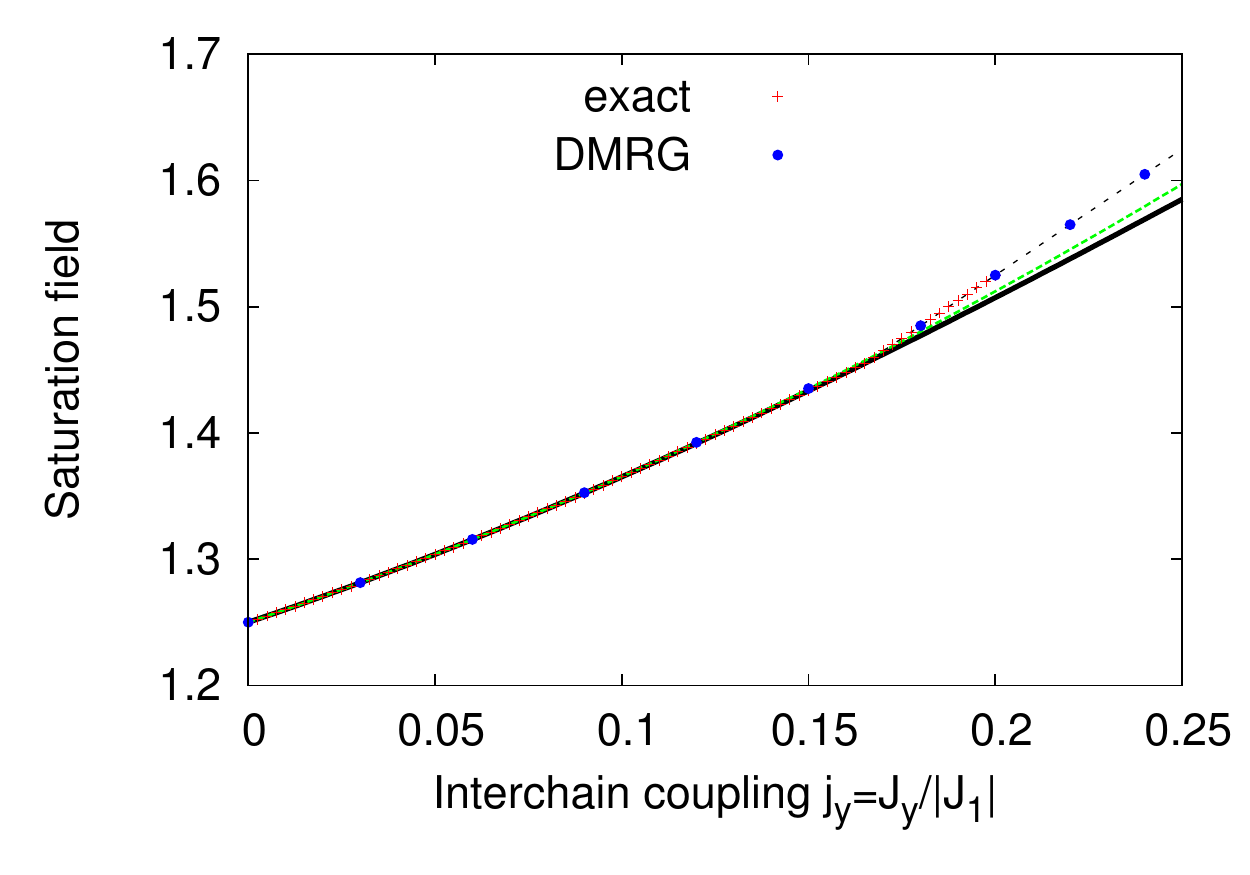} 
\caption{(Color online) The saturation field 
$h_{s,2}(\alpha,j_{y})=\mu \mathcal{H}_{s}/|J_{1}|$
for a 2D array of chains ($J_y\equiv J_{\perp}$) for $\alpha=1$,$\Delta _1 = 1$. 
Black solid line: the
result of analytic Eq.(\ref{eq:hs2bet4}), green short-dashed line:
the result of the expansion (\ref{eq:hs2bet4}) up to second order
(i.e.\ $\zeta_{4}$ is neglected),  black dashed line: 
the field of the
1-magnon instability $h_{s,1}$ (\ref{eq:hs1}). 
Points: DMRG-data and data from the numerical solution of Eq.\ (\ref{eq:pole}).}
\label{fig4} 
\end{figure}

Assuming $\zeta\ll1$, we rewrite Eq.\ (\ref{eq:polepi}) in the
form\[
\frac{1}{N_{\perp}}\sum_{q_{y},q_{z}}\sum_{m=0}^{\infty}\frac{G^{(m)}}{m!}\left(\zeta-e_{\mathbf{q}}\right)^{m}=-\frac{\alpha}{\Delta_{1}},\]
 where $e_{\mathbf{q}}\equiv N_{ic}J_{\perp}\gamma_{\mathbf{q}}/J2$,
\[
G^{(m)}\equiv\left(\frac{\partial}{\partial z}\right)^{m}G(z)\mid_{z=z_{b1}}.\]
 Note that $G(z_{b1})=-\alpha/\Delta_{1}^{z}$, and keeping only
terms with $m\leq4$, we obtain the equation\begin{eqnarray}
\zeta G^{\prime}+\frac{1}{2}\left(\zeta^{2}+\overline{e_{\mathbf{q}}^{2}}\right)G^{\prime\prime}\nonumber \\
+\frac{1}{6}\left(\zeta^{3}+3\zeta\overline{e_{\mathbf{q}}^{2}}\right)G^{\prime\prime\prime}\nonumber \\
+\frac{1}{24}\left(\zeta^{4}+6\zeta^{2}\overline{e_{\mathbf{q}}^{2}}+\overline{e_{\mathbf{q}}^{4}}\right)G^{IV} & = & 0,\label{eq:dz4eq}\end{eqnarray}
 where 
 \[
\overline{e_{\mathbf{q}}^{m}}\equiv\frac{1}{N_{\perp}}\sum_{q_{y},q_{z}}e_{\mathbf{q}}^{m},
\]
 and we have taken into account that $\overline{e_{\mathbf{q}}}=\overline{e_{\mathbf{q}}^{3}}=0$.
The direct calculation yields 
$\overline{e_{\mathbf{q}}^{2}}=N_{ic}\left(J_{\perp}/J_{2}\right)^{2}$,
and $\overline{e_{\mathbf{q}}^{4}}=6\left(J_{\perp}/J_{2}\right)^{4}$
($36\left(J_{\perp}/J_{2}\right)^{4}$) for 2D(3D) respectively;
\begin{eqnarray}
G^{\prime} & = & G^{2}\left[\tau^{\prime}-1\right],\label{eq:Gp}\\
G^{\prime\prime} & = & 2G^{3}\left[\tau^{\prime}-1\right]^{2}+G^{2}\tau^{\prime\prime},\label{eq:G2p}\\
G^{\prime\prime\prime} & = & 6G^{4}\left[\tau^{\prime}-1\right]^{3}\nonumber \\
 & + & 6G^{3}\left[\tau^{\prime}-1\right]\tau^{\prime\prime}+G^{2}\tau^{\prime\prime\prime},\label{eq:G3p}\\
G^{IV} & = & 24G^{5}\left[\tau^{\prime}-1\right]^{4}+36G^{4}\left[\tau^{\prime}-1\right]^{2}\tau^{\prime\prime}\nonumber \\
 & + & 6G^{3}\left(\tau^{\prime\prime}\right)^{2}+8G^{3}\left[\tau^{\prime}-1\right]\tau^{\prime\prime\prime}+G^{2}\tau^{IV},\label{eq:G4p}\\
\tau^{\prime} & = & -\frac{\alpha^{2}}{\Delta_{1}\left(\Delta_{1}+2\alpha\right)},\nonumber \\
\tau^{\prime\prime} & = & -2\left[\frac{\alpha\left(\Delta_{1}+\alpha\right)}{\Delta_{1}\left(\Delta_{1}+2\alpha\right)}\right]^{3},\nonumber \\
\tau^{\prime\prime\prime} & = & -6\left[\frac{\alpha\left(\Delta_{1}+\alpha\right)}{\Delta_{1}\left(\Delta_{1}+
2\alpha\right)}\right]^{4}\frac{\Delta_{1}^{2}+2\Delta_{1}\alpha+2\alpha^{2}}{\Delta_{1}\left(\Delta_{1}+2\alpha\right)},\nonumber \\
\tau^{IV} & = & -24\frac{\alpha^{5}\left(\Delta_{1}+\alpha\right)^{5}F}{\left[\Delta_{1}\left(\Delta_{1}+2\alpha\right)\right]^{7}},\nonumber \\
F & \equiv & \Delta_{1}^{4}+4\Delta_{1}^{3}\alpha+9\Delta_{1}^{2}\alpha^{2}+10\Delta_{1}\alpha^{3}+5\alpha^{4}. \nonumber 
\end{eqnarray}
 Substituting the expansion
 \begin{equation}
\zeta=\zeta_{1}j_{ic}+\zeta_{2}j_{ic}^{2}+\zeta_{3}j_{ic}^{3}+\zeta_{4}j_{ic}^{4},\label{eq:dzeta}
\end{equation}
 ($j_{ic}\equiv J_{\perp}/|J_{1}|$) into (\ref{eq:dz4eq}), we obtain
$\zeta_{1}=\zeta_{3}=0$, and\begin{eqnarray}
\hspace{-0.25cm}\zeta_{2} & = & -\frac{N_{ic}G^{\prime\prime}}{2\alpha^{2}G^{\prime}} \  , \label{eq:dz2gen}\\
\hspace{-0.25cm}  & = & \hspace{-0.1cm} -\frac{N_{ic}\left(\Delta_{1}+\alpha\right)}{\alpha\left[\Delta_{1}\left(\Delta_{1}+2\alpha\right)\right]^{2}}\left[\Delta_{1}^{2}+3\Delta_{1}\alpha+3\alpha^{2}\right],\label{eq:dz2}\\
\hspace{-0.25cm} \zeta_{4} & = & -\frac{1}{G^{\prime}}\left[\frac{G^{\prime\prime}}{2}\zeta_{2}^{2}-\frac{N_{ic}G^{\prime\prime\prime}}{2\alpha^{2}}\zeta_{2}+\frac{G^{IV}}{24\beta^{4}}\overline{e_{\mathbf{q}}^{4}}\right].\label{eq:dz4}\end{eqnarray}
 At the saturation field, the $\omega_{b}$ in the right-hand side
of Eq.\ (\ref{eq:wb}) vanishes, and we obtain
\begin{eqnarray}
\hspace{-0.4cm} h_{s,2} & = & h_{s,2}^{\rm 1D}+\frac{N_{\rm ic}}{2}j_{\rm ic}\Delta_{\perp}-
\frac{\alpha}{2}\left(\zeta_{2}j_{\rm ic}^{2}+\zeta_{4}j_{\rm ic}^{4}\right) \ , \ \label{eq:hs2bet4}\\
h_{s,2}^{\rm 1D} & = & -\Delta_{1}+\alpha\Delta_{2}-\frac{\alpha}{2}z_{\rm b1} \quad ,\nonumber \end{eqnarray}
 where $h_{s}\equiv \mu \mathcal{H}_{s}/|J_{1}|$. Eq.\ (\ref{eq:hs2bet4})
coincides with 
Eq.\ (3) of the main text with $\eta_{i}=-\alpha\zeta_{i}/2$.
Its validity is demonstrated in Fig.\ \ref{fig4}. 
As an example, we have chosen the 2D case and $\alpha =1$, i.e.\ 
the optimal region for the 
existence of the nematic phase, where $j_{\rm cr} \approx 0.167$.
\begin{figure}
\includegraphics[width=.45\columnwidth]{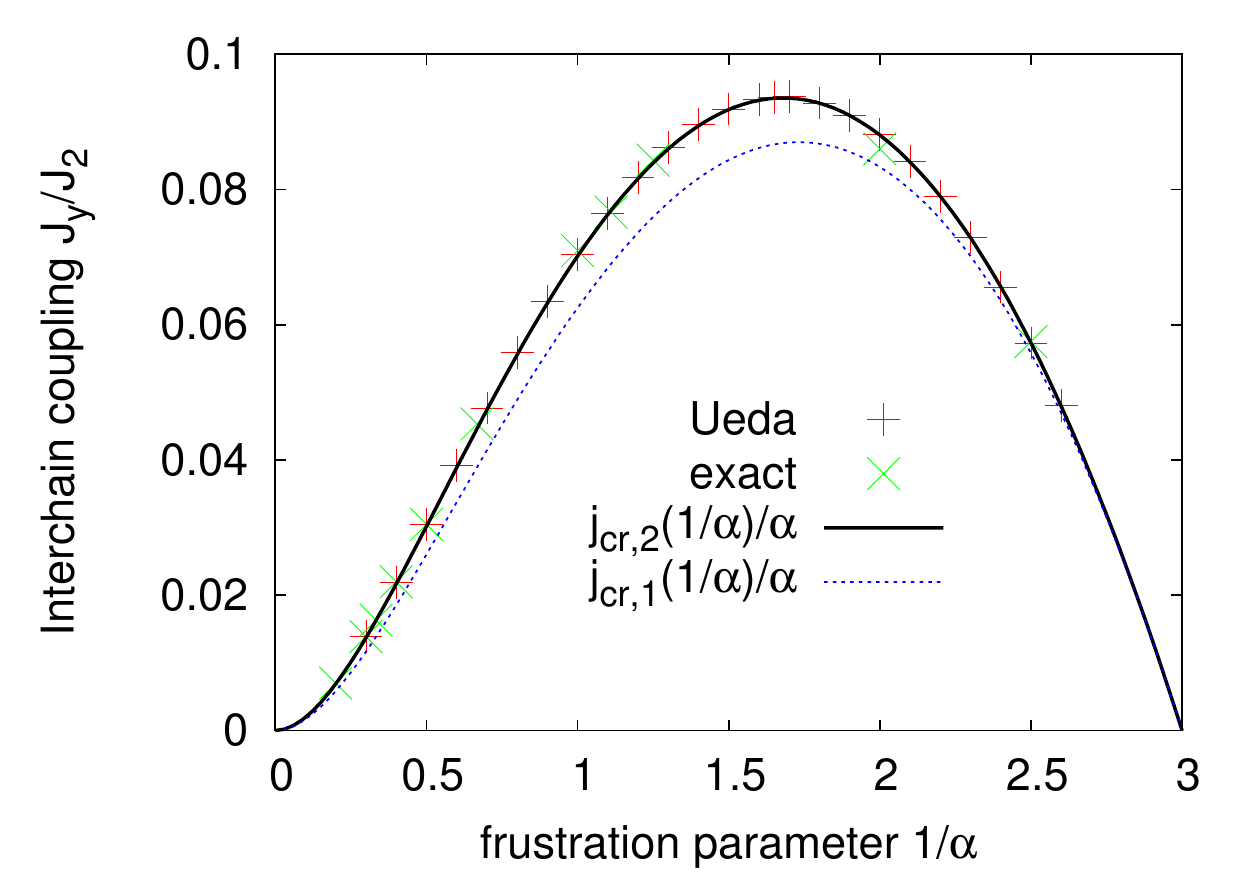} 
\caption{Boundary between the 1- and 2- magnon phases for the 3D case 
($J_y\equiv J_{\perp}$). Points: the numerical results from this work 
and from Ref.\ \onlinecite{Ueda09}}
\label{fig6} 
\end{figure}
\begin{figure}[b]
\includegraphics[width=.45\columnwidth]{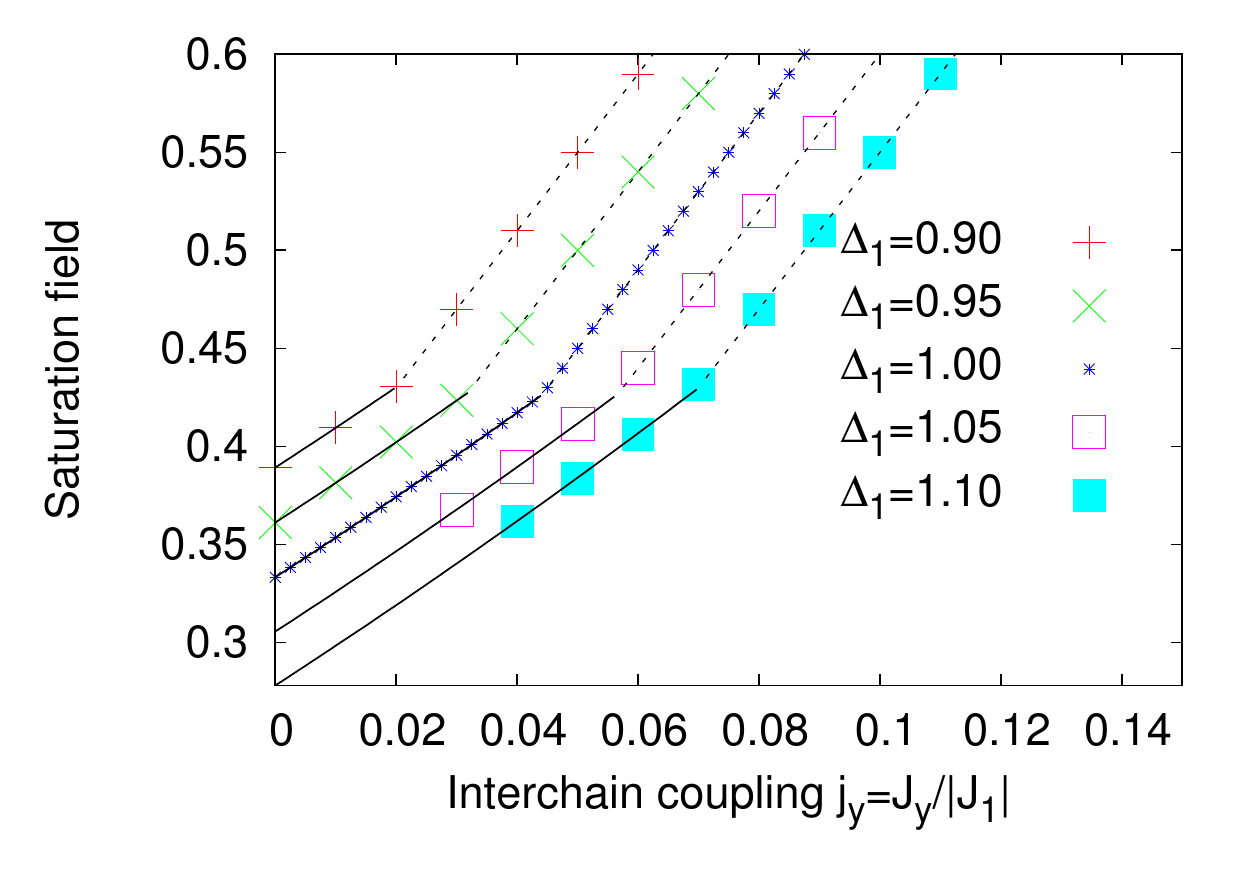} 
\caption{The saturation field $h_{s,2}$ for the 3D case for 
$\alpha=0.5$ ($J_y\equiv J_{\perp}$).
The easy-axis anisotropy of the NN coupling
is taken into account. The meaning of the lines is the same as in Fig.\ \ref{fig4}.
Points: DMRG-data ($\Delta _1 \neq 1$), and the data
from a numerical solution of Eq.\ (\ref{eq:pole}) ($\Delta _1 = 1$)}.
\label{fig5} 
\end{figure}
For small $j_{\perp}=J_{\perp}/|J_1|$ the second order expansion 
reproduces well the DMRG data which coincide with the results from a 
numerical solution of Eq.\ (\ref{eq:pole}).
Naturally, for larger interchain coupling $j_{\perp} > 0.15$ the fourth
order expansion is needed.

The boundary between the 1-magnon and the 2-magnon phases is 
obtained by solving
the equation 
$h_{s,2}(j_{\rm cr})=h_{s,1}(j_{\rm cr, })$ 
for the critical IC $j_{\rm cr}$.
If one retains only the linear term in the expansion in powers of the IC 
given by (\ref{eq:hs2bet4}),
we obtain (cf.\ Eq.\ (51) in Ref.\ \onlinecite{Syromyatnikov12})
\begin{equation}
|j_{\rm cr,1}|=\frac{4\alpha\Delta_{1}^{2}-\Delta_{1}-\alpha}
{4N_{\rm ic}\alpha\left(\Delta_{1}+\alpha\right)}.
\label{eq:bet1}
\end{equation}
 This approximation demonstrates the qualitative 
 behaviour of $j_{\rm cr}$ as a
function of the anisotropy and the frustration parameters
$\Delta _1$ and $\alpha$, respectively. Practically, a fully quantitative
agreement with our numerical data is achieved, if we account also for the 
quadratic term in Eq.\ (\ref{eq:hs2bet4}) 
\begin{equation}
|j_{cr,2}|=\frac{1}{2\alpha\zeta_{2}}\left(-N_{ic}+\sqrt{N_{ic}^{2}+
4N_{ic}\alpha\zeta_{2}|j_{cr,1}|}\right).
\label{eq:bet2}
\end{equation}
It is convenient to normalize the couplings on $J_{2}>0$, 
and introduce $\kappa\equiv1/\alpha$, which measures
the attraction provided by the FM $J_1$. Using the same normalization for
the IC, too, we write $y\equiv J_{\perp}/J_{2}=j_{ic}/\alpha$.
Then the Eqs.\  (\ref{eq:dz2}), (\ref{eq:bet1}), and (\ref{eq:bet2})
may be rewritten as 
\begin{eqnarray}
\hspace{-0.5cm}\zeta_{2} & = & \hspace{-0.1cm}-\frac{N_{ic}\left(\kappa\Delta_{1}+1\right)}{\left[\Delta_{1}\left(\kappa\Delta_{1}+
2\right)\right]^{2}}\left[\kappa2\Delta_{1}^{2}+3\Delta_{1}\kappa+3\right],\label{eq:dzkap}\\
\hspace{-0.5cm}|y_{cr,1}| & = &  \frac{\kappa^{2}\left(4\Delta_{1}^{2}-\kappa\Delta_{1}-1\right)}{4N_{ic}\left(\kappa\Delta_{1}+1\right)},\label{eq:y1}\\
\hspace{-0.5cm}|y_{cr,2}| & = & \frac{\kappa ^2}{2\zeta_{2}}\left(-N_{ic}+\sqrt{N_{ic}^{2}+\frac{4N_{ic}\zeta_{2}|y_{cr,1}|}{\kappa ^2}}\right). 
\label{eq:y2}\end{eqnarray}

A comparison of the results of the approximate analytic Eqs.\ (\ref{eq:y1}) 
and (\ref{eq:y2}) with the numerical data is shown in Fig.\ \ref{fig6}. Note the high
accuracy achieved already in the second order of the IC in 
Eq.\ (\ref{eq:hs2bet4}).
Finally, an example of the saturation field dependence on the anisotropy 
parameter is shown in Fig.\ \ref{fig5}.


\end{document}